\documentclass[usenatbib]{mnras}

\usepackage[utf8]{inputenc}
\usepackage[T1]{fontenc}
\usepackage{ae,aecompl}
\usepackage{geometry}
\usepackage{pdflscape}

\usepackage{graphicx}
\usepackage{float}
\usepackage{amssymb}
\usepackage{amsmath}
\usepackage{relsize}
\usepackage{cleveref}
\crefformat{section}{§#2#1#3}
\crefname{section}{§}{§§}
\Crefname{section}{§}{§§}

\def\beq{\begin{eqnarray}}
\def\eeq{\end{eqnarray}}

\def\ngal{n_{\mathrm{gal}}}
\def\ncluster{n_{c}}
\def\siglogM{\sigma_{\log M}}
\def\delgam{\Delta \gamma}
\def\Mmin{M_{\mathrm{min}}}
\def\satfrac{\ln \left( M_1 / M_{\mathrm{min}}\right)}
\def\M0oM1{\ln \left ( \frac{M_0}{M_1} \right )}
\def\textM0oM1{\ln \left( M_0 / M_1 \right)}
\def\siglnMc{\sigma_{\ln M_c}}
\def\Qenv{Q_{\mathrm{env}}}
\def\Mpch{h^{-1} \; \mathrm{Mpc}}
\def\MpchCubed{\mathrm{h^{-3}} \; \mathrm{Mpc}^3}
\def\invMpchCubed{\mathrm{h^3}\;\mathrm{Mpc^{-3}}}
\def\Msun{\mathrm{M_{\odot}}}
\def\pimax{\Pi_{\mathrm{max}}}
\def\Sigsrc{\Sigma_{\mathrm{src}}}
\def\Sigcrit{\Sigma_{\mathrm{crit}}}
\def\rmin{r_{\mathrm{min}}}
\def\rmax{r_{\mathrm{min}}}

\def\nfid{n_{\mathrm{fid}}}

\title[Cluster Forecasting]{Cosmology with Stacked Cluster Weak Lensing and Cluster-Galaxy Cross-Correlations}

\author[A. N. Salcedo et al.]{Andr\'{e}s N. Salcedo$^{1}$\thanks{E-mail: salcedo.11@osu.edu},
Benjamin D. Wibking$^{1}$,
David H. Weinberg$^{1}$,
Hao-Yi Wu$^{1}$, \newauthor
Douglas Ferrer$^{2}$,
Daniel Eisenstein$^{2}$,
and Philip Pinto$^{3}$
\\
$^{1}$ Department of Astronomy and Center for Cosmology and AstroParticle Physics, The Ohio State University, Columbus, OH 43210, USA \\
$^{2}$ Harvard-Smithsonian Center for Astropyhsics, 60 Garden St., MS-10, Cambridge, MA 02138 \\
$^{3}$ Steward Observatory, University of Arizona, 933 N. Cherry Ave., Tucson, AZ 85121}

\date{Accepted XXX. Received YYY; in original form ZZZ}

\pubyear{2019}

\begin{document}
\label{firstpage}
\pagerange{\pageref{firstpage}--\pageref{lastpage}}
\maketitle

\begin{abstract}
Cluster weak lensing is a sensitive probe of cosmology, particularly the amplitude of matter clustering $\sigma_8$ and matter density parameter $\Omega_m$. The main nuisance parameter in a cluster weak lensing cosmological analysis is the scatter between the true halo mass and the relevant cluster observable, denoted $\siglnMc$. We show that combining the cluster weak lensing observable $\Delta \Sigma$ with the projected cluster-galaxy cross-correlation function $w_{p,cg}$ and galaxy auto-correlation function $w_{p,gg}$ can break the degeneracy between $\sigma_8$ and $\siglnMc$ to achieve tight, percent-level constraints on $\sigma_8$. Using a grid of cosmological N-body simulations, we compute derivatives of $\Delta \Sigma$, $w_{p,cg}$, and $w_{p,gg}$ with respect to $\sigma_8$, $\Omega_m$, $\siglnMc$ and halo occupation distribution (HOD) parameters describing the galaxy population. We also compute covariance matrices motivated by the properties of the Dark Energy Suvery (DES) cluster and weak lensing survey and the BOSS CMASS galaxy redshift survey. For our fiducial scenario combining $\Delta \Sigma$, $w_{p,cg}$, and $w_{p,gg}$ measured over $0.3-30.0 \; \Mpch$, for clusters at $z=0.35-0.55$ above a mass threshold $M_c\approx 2\times 10^{14} \; h^{-1} \; \Msun$, we forecast a $1.4\%$ constraint on $\sigma_8$ while marginalizing over $\siglnMc$ and all HOD parameters. Reducing the mass threshold to $1\times 10^{14} \; h^{-1} \; \Msun$ and adding a $z=0.15-0.35$ redshift bin sharpens this constraint to $0.8\%$. The small scale $(r_p < 3.0 \; \Mpch)$ ``mass function'' and large scale $(r_p > 3.0 \; \Mpch)$ ``halo-mass cross-correlation'' regimes of $\Delta \Sigma$ have comparable constraining power, allowing internal consistency tests from such an analysis.
\end{abstract}


\section{Introduction}

The abundance of rich galaxy clusters as a function of mass provides a sensitive probe of the amplitude of matter clustering $\sigma_8$ and the matter density parameter $\Omega_m$ \citep[][for recent reviews see \citet{Allen_Evrard_Mantz_ARA_2011} and chapter 6 of \citet{Weinberg_PhR_2013}, hereafter WMEHRR]{Evrard_1989,White_Efstathiou_Frenk_1993}. Although this approach is usually applied on scales of the cluster virial radius, large scale cluster-mass correlations probed by weak gravitational lensing also constrain $\sigma_8$ and $\Omega_m$ \citep{Zu_et_al_2014}. When combined with a model of non-linear galaxy bias, the mass-to-light or mass-to-number ratios of clusters can also constrain $\sigma_8$ and $\Omega_m$, by a conceptually distinct route with different sensitivity from the halo mass function alone \citep{vdBosch_Mo_Yang_2003, Tinker_et_al_2005, Vale_Ostriker_2006, Tinker_et_al_2012}. As strongly clustered tracers that can be observed over large volumes, galaxy clusters also probe the amplitude and shape of the matter power spectrum $P(k)$ through their auto-correlation function \citep[e.g.][]{Bahcall_Soneira_1984, Croft_Efstathiou_1994, Croft_et_al_1997, Bahcall_et_al_2003, Estrada_et_al_2009} or their cross-correlation with galaxies \citep{Croft_Dalton_Efstathiou_1999, Sanchez_et_al_2005, Paech_et_al_2017}.

In this paper we investigate the constraints on $\sigma_8$ and $\Omega_m$ that can be obtained by combining cluster excess surface density profiles $\Delta \Sigma (r_p)$ measured by weak lensing with the projected cluster-galaxy cross-correlation function $w_{p,cg}(r_p)$ and galaxy auto-correlation function $w_{p,gg}(r_p)$ (see \cref{sec:derivs} for definition). Cluster mass is not directly observable, but many observable properties of clusters are correlated with mass, such as galaxy richness, total stellar mass, galaxy velocity dispersion, X-ray luminosity, X-ray temperature, X-ray inferred gas mass, or integrated Sunyaev-Zeldovich decrement \citep[][SZ]{SZ_1970}. Weak lensing plays an indispensable role in cluster cosmology because it allows calibration of the mean mass-observable relation with the minimum sensitivity to uncertainties in baryonic physics \citep{Sheldon_et_al_2009B, Rozo_et_al_2010, vdLinden_et_al_2014,Melchior_DES_2017,McClintock_DES_2019}. For the approach described in this paper, we have in mind wide area, deep imaging surveys such as the Dark Energy Survey \citep[DES;][]{DES_2005}, the Subaru Hyper-Suprime Camera survey \citep[HSC;][]{HSC_Aihara_2018a}, and in the future, surveys by the Large Synoptic Survey Telescope \citep[LSST;][]{LSST_Science_Book_v2_2009}, the Euclid mission \citep[]{Laureijs_Euclid_2011}, and the Wide Field Infrared Survey Telescope \citep[WFIRST;][]{Dore_WFIRST_2018}. These surveys allow high-precision measurements of weak lensing profiles, for clusters identified from the survey galaxy distribution or from external X-ray or SZ data sets. Galaxies with photometric redshifts from the surveys can be used to measure $w_{p,cg}$ and $w_{p,gg}$. 

Cluster cosmological studies frequently focus on inferring the halo mass function from cluster counts as a function of a mass proxy observable \citep[e.g.][]{Vikhlinin_et_al_2009, Mantz_et_al_2010, Benson_SPT_et_al_2013,Reichardt_SPT_et_al_2013} or on directly forward modeling the counts of clusters as a function of these mass proxies \citep[e.g.][]{Rozo_et_al_2010,Constanzi_et_al_2018}. With a large weak lensing survey one can treat cluster cosmology as more closely analagous to galaxy-galaxy lensing, measuring the space density and mean $\Delta \Sigma (r_p)$ profile of all clusters above a threshold in the observable. This philosophy, similar to that advocated by \citet{Zu_et_al_2014} and WMEHRR, is the one we adopt here. The most important astrophysical nuisance parameter in such a study is the rms fractional scatter between the true halo mass and cluster observable, denoted $\siglnMc$ in this paper. For a cluster sample defined by a threshold, one only needs to know $\siglnMc$ at the threshold, while an analysis that uses bins of observables requires $\siglnMc$ at all bin boundaries. \citet{Oguri_Takada_2011} and WMEHRR show that with good knowledge of $\siglnMc$ the cosmological constraints expected from cluster weak lensing are competitive with, and complementary to, those expected from cosmic shear analysis of the same weak lensing data set.
 
 There are several ways to think about the potential gains from combining $\Delta \Sigma$, $w_{p,cg}$, and $w_{p,gg}$. First, one can view $w_{p,cg}$ and $w_{p,gg}$ as observables to constrain $\siglnMc$ and thus break degeneracy with cosmological parameters. Second, on large scales where linear theory and scale-independent bias should be good approximations, we expect (for fixed $\Omega_m$) $\Delta \Sigma \propto b_c \sigma_8^2$, $w_{p,cg} \propto b_c b_g \sigma_8^2$, and $w_{p,gg} \propto b_g^2 \sigma_8^2$. Three observables are sufficient to determine the three unknowns. Finally, on small scales our three-observable approach resembles the mass-to-number ratio method of \citet{Tinker_et_al_2012}, as $\Delta \Sigma(r_p)$ and $w_{p,cg}(r_p)$ provide projected cluster mass and number density profiles and $w_{p,gg}(r_p)$ provides the galaxy clustering constraints on galaxy halo occupation. These three interpretations are not mutually exclusive and not fully separable, though we attempt (in \cref{sec:forecasts}) to disentangle the strands of information in our approach by examining the contribution from different observables on different scales. We find in our forecasts that the constraints on $\sigma_8$ from the combination of all three observables are far tighter than those from any pairwise combination of them.
 
 To use galaxy clustering observables down to sub-$\mathrm{Mpc}$ scales we need a fully non-linear model of the relation between galaxies and mass. For this purpose we use the halo occupation distribution \citep[HOD;][]{Berlind_2002} and marginalize over HOD parameters when constraining cosmological parameters as advocated by \citet{Zheng_Weinberg_2007}. We follow the approach of \citet{Wibking_et_al_2019A} to obtain accurate predictions in the non-linear regime by populating N-body halos from the {\sc{AbacusCosmos}} suite of cosmological simulations \citep{Metchnik_2009,Garrison_et_al_2018}. Like \citet{Wibking_et_al_2019A}, we include an extended HOD parameter that allows the halo occupation to vary with large scale environment, to represent the possible effects of galaxy assembly bias \citep{Hearin_et_al_2016,Zentner_et_al_2019}. A similar approach to emulating galaxy clustering with a grid of populated N-body simulations is presented by \citet{Zhai_AemulusII_2019}.
 
 The next section describes in detail our numerical simulation suite, HOD modeling methodology, and model of the cluster mass-observable relation. Section \ref{sec:gal_and_cluster} defines our clustering and weak lensing statistics and derives their sensitivity to HOD and cosmological parameters, with Figures \ref{fig:cg_frac_diffs} and \ref{fig:gg_frac_diffs} as the key summary plots. Section \ref{sec:cov} describes how we estimate the error covariance matrices of $w_{p,cg}$, $w_{p,gg}$, and $\Delta \Sigma$ for our fiducial forecast, which is based loosely on the properties of DES. We present our main results in \cref{sec:forecasts}, combining the derivatives of \cref{sec:derivs}  with the covariances of \cref{sec:cov} to forecast the $\sigma_8$ and $\Omega_m$ constraints that can be obtained from various combinations of the three observables on small ($r_p = 0.3 - 3.0 \; \Mpch$) and large ($r_p = 3.0 - 30.0 \; \Mpch$) scales. Table \ref{table:uncertainties} and figure \ref{fig:fid_forecast} contain the key quantitative results. We summarize our findings in \cref{sec:conc} and identify directions for future work. 
  
\section{Creating Simulated Galaxy and Cluster Populations}
\label{sec:gal_and_cluster}

\subsection{Numerical Simulations}
\label{sec:sims}

\begin{table}
   \centering
   \begin{tabular}{cccccccc}
      \hline
      Name  & $h$ & $N_{eff}$ & $\Omega_{\Lambda}$ & $\Omega_m$ & $n_s$ & $\sigma_8$ & $w_0$ \\
 	\hline
      Emu00  & 0.673  & 3.04 & 0.686 & 0.314 & 0.965 & 0.83 &  -1.0 \\
      Emu01  &   .    &  .   &  .    &  .    &  .    & 0.78 &    .  \\
      Emu02  &   .    &  .   &  .    &  .    &  .    & 0.88 &    .  \\
      Emu03  & 0.643  &  .   & 0.656 & 0.344 &  .    & 0.83 &    .  \\
      Emu04  & 0.703  &  .   & 0.712 & 0.288 &  .    &  .   &    .  \\
	\hline
   \end{tabular}
   \caption{Cosmological grid in $\sigma_8$ and $\Omega_m$ used in this analysis. We compute derivatives of observables with respect to cosmological parameters using a single realization of each model with matched Fourier phases. We use 20 realizations of the Emu00 cosmology to compute derivatives with respect to HOD parameters and covariance matrices. Each realization models an $1100.0 \; \Mpch$ cube with $1440^3$ particles.}
   \label{table:sims}
\end{table}

In this paper we use a five simulation grid in cosmology ($\Omega_m$, $\sigma_8$) centred on a flat $\Lambda$CDM cosmological model based on the Planck \citep{Planck_2016} satellite's measurements; $\Omega_m = 0.314$, $\Omega_\Lambda = 0.686$, $h = 0.673$, $\sigma_8 = 0.83$, $n_s = 0.965$, and $N_{\mathrm{eff}} = 3.04$. The values of our steps up and down in $\Omega_m$ and $\sigma_8$ are shown in Table \ref{table:sims}; when varying $\Omega_m$ we hold $\Omega_m h^2$ and $\sigma_8$ fixed. All five boxes are periodic cubes with side length $L = 1100.0 \; \Mpch$ that contain $N_p = 1440^3$ particles with a Plummer gravitational softening length of $\epsilon_g = 62.5\; \mathrm{h^{-1}} \; \mathrm{kpc}$. These boxes are all evolved from one fixed set of phases. We also use an additional 20 runs of the fiducial cosmology with different initial phases to measure box-to-box variance and numerically calculate covariance matrices for our observables.

Our simulations are run using the {\sc{abacus}} \citep{Metchnik_2009,Garrison_et_al_2018} cosmological N-body code. {\sc{abacus}} attains both speed and accuracy by utilizing novel computational techniques and high performance hardware such as GPUs and RAID disk arrays. Force computations are split into near-field and far-field components. Near-field forces are computed directly, while the far field component is calculated from the multipole moments of particles in the cells \citep{Metchnik_2009}. To determine the initial conditions, CAMB \citep{Lewis_2011} is used to compute an input $z = 0$ power spectrum of density fluctuations, which is then scaled back to $z = 49$ via a ratio of growth factors. An initial density field at $z = 49$ is then generated with initial particle positions and velocities using {\sc{abacus}}' second order Lagrangian perturbation theory (2LPT) implementation with rescaling \citep{Garrison_2016}. Since 2LPT accounts for early non-linear gravitational evolution, it is more accurate than the Zeldovich approximation \citep{Zeldovich_1970_ZelApprox}, particularly in the case of the rarest high density peaks. {\sc{abacus}} improves upon standard 2LPT by rescaling growing modes near $k_\mathrm{Nyquist}$ that are supressed due to the effect of treating dark matter as discrete macroparticles. Once the initial conditions are specified, gravitational evolution is performed using {\sc{abacus}} and particle snapshots are saved at multiple redshifts. Most of our results and figures are based on the $z = 0.5$ snapshots of these simulations. In the forecast section we consider the impact of adding a second lower redshift cluster bin, which we model with the $z = 0.3$ output.

\subsection{Halo Identification}
\label{sec:hfind}

We use the software package {\sc{rockstar}} version 0.99.9-RC3+ \citep{Behroozi_2013} to identify haloes from the particle snapshots. However we use strict (i.e., without unbinding) spherical overdensity (SO) halo masses around the halo centres identified by {\sc{rockstar}}, rather than the default phase-space FOF-like masses output by {\sc{rockstar}}. For finding haloes {\sc{rockstar}} uses a primary definition set to the virial mass of \citet{Bryan_1998}. However, after identification, we adopt the $M_{200b}$ mass definition, i.e.,  the mass enclosed by a spherical overdensity of 200 times the mean matter density at a given redshift and cosmology. Distinct haloes identified with the $M_{vir}$ definition are not reclassified as subhalos under the $M_{200b}$ definition; such reclassification would affect a negligible fraction of halos. We identify halos above 20 particles, and we only use distinct halos (not subhalos) when creating galaxy populations.

\subsection{HOD Modeling}
\label{sec:HOD}

\begin{table}
   \centering
    \begin{tabular}{cc p{3.2cm}}      
    \hline
      Parameter  & Fiducial Value & Description \\
 	\hline
      $\ngal \times 10^4$ & $2.18 \; \invMpchCubed$ & galaxy number density \\
      $\siglogM$ & $0.6$ & width of central occupation cutoff \\
      $\alpha$ & $1.6$ & slope of satellite occupation power law \\
      $\ln \left( \frac{M_1}{M_{\mathrm{min}}} \right)$ & $0.9$ & satellite fraction parameter\\
      $\M0oM1$ & $-13.7$ & satellite cutoff parameter\\
      $\Qenv$ & $0.0$ & environmental dependence of galaxy occupation parameter \\
      $\delgam$ & $0.0$ & galaxy-deviation from NFW parameter \\\
      $\ncluster \times 10^6$ & $3.228 \; \invMpchCubed$ & cluster number density \\
      $\siglnMc$ & $0.4$ & cluster mass-observable scatter \\
      $\Omega_m$ & $0.314$ & cosmological matter density \\
      $\sigma_8$ & $0.83$ & power-spectrum amplitude \\
	\hline
   \end{tabular}
   \caption{Fiducial Model Parameters (HOD and Cosmological).}
\label{table:params}
\end{table}

We populate our simulated haloes with galaxies according to a halo occupation distribution (HOD) framework \citep[e.g.][]{Jing_Mo_Borner_1998, Benson_et_al_2000, Peacock_Smith_2000, Seljak_2000,Scoccimarro_et_al_2001, Berlind_2002, Zheng_et_al_2005, Zheng_2009, Zehavi_et_al_2011, Coupon_et_al_2012, Guo_2014, Zu_Mandelbaum_12-2015, Zehavi_et_al_2018}. In this framework it is helpful to separate the hosted galaxies into satellites and centrals \citep{Guzik_Seljak_2002, Kravtsov_et_al_2004}. According to this prescription haloes tend to host exactly one central above some mass, and satellite occupation is an increasing power law in mass. We parametrize the mean occupation number of our haloes as

\begin{align}
\left \langle N_{\mathrm{cen}}(M) \right \rangle &= \frac{1}{2} \left [ 1 + \mathrm{erf} \left ( \frac{\log M - \log M_{\mathrm{min}}}{\siglogM}\right ) \right ], \label{eq:cen_HOD} \\ 
\left \langle N_{\mathrm{sat}}(M) \right \rangle &= \left \langle N_{\mathrm{cen}} (M) \right \rangle \left ( \frac{M - M_0}{M_1} \right )^{\alpha}. \label{eq:sat_HOD}
\end{align}
The actual number of centrals placed into a given halo is either zero or one and is determined randomly given the mean central occupation. The number of satellites placed into a halo is sampled from a Poisson distribution centred at the mean satellite occupation.

There are five free parameters in this prescription\footnote{Note that we use $\log = \log_{10}$ and $\ln = \log_{e}$ throughout.}. The parameter $M_{\mathrm{min}}$ sets the mass scale at which haloes start hosting a central, i.e. $\left \langle N_{\mathrm{cen}}(M_{\mathrm{min}}) \right \rangle = 0.5$. The sharpness of the transition from $\left \langle N_{\mathrm{cen}}(M) \right \rangle = 0.0$ to $\left \langle N_{\mathrm{cen}}(M) \right \rangle = 1.0$ is determined by the parameter $\siglogM$. This transition is a step function smoothed to a width of $\siglogM$ to model the scatter between halo mass and central galaxy luminosity. The parameters $M_1$ and $M_0$ are the satellite normalization scale and satellite cut-off scale respectively and satisfy $\left \langle N_{\mathrm{sat}}(M_1 + M_0) = 1.0 \right \rangle$. In practice we find that our fiducial value of $M_0$ (based on \citealt{Guo_2014}) is so small compared to typical halo masses in our simulations that $M_0$ has negligible effect on clustering measurements. Finally the parameter $\alpha$ determines the slope of the satellite occupation. This parameterization is the same as that of \citet{Zheng_et_al_2005}.

We place central galaxies at the centre of their host haloes. Satellites are distributed according to a generalized Navarro-Frenk-White \citep[][NFW]{NFW_1997} profile,

\beq
\rho_{\mathrm{gal}} (r) = \rho_{m} (r | c_{\mathrm{vir}}) r^{\delgam},
\label{eq:delgam}
\eeq
parametrized by halo concentration $c_{\mathrm{vir}} = r_{h} / r_{s}$. Previous studies \citep[e.g.][]{Power_et_al_2002, Navarro_et_al_2004,Springel_et_al_2008, Diemer_Kravtsov_2015} have shown that certain halo properties, including $c_{\mathrm{vir}}$, require up to and beyond 1000 particles to converge. Thus, because of our mass and force resolution we choose to assign values of $c_{\mathrm{vir}}$ to haloes using fits to the halo concentration-mass relationship from \citet{Correa_2015}, which were calibrated with simulations at significantly better resolution:

\begin{align}
\log c_{\mathrm{vir}} &= \alpha + \beta \log (M / \Msun) \left [ 1 + \gamma \log^2 \left (M / \Msun \right ) \right ]  \\
\alpha &= 1.62774 - 0.2458 ( 1 + z) + 0.01716 (1 + z)^2 \nonumber \\
\beta &= 1.66079 + 0.00359(1 + z) - 1.6901 (1 + z)^{0.00417} \nonumber \\
\gamma &=  -0.02049 + 0.0253 (1 + z)^{-0.1044} . \nonumber
\end{align}
We further approximately rescale from the $M_{200c}$ mass definition used in \citet{Correa_2015} to $M_{200b}$ by multiplying the concentration by $\sqrt{2}$ \citep{Hu_Kravstov_2003}.

Depending on the value of $\delgam$, the galaxy profile can deviate from that of the matter, which follows a NFW profile parametrized by halo concentration $c_{\mathrm{vir}}$, while still inheriting the geometry of the halo. The $\delgam$ parameter also has the advantage of adding flexibility to compensate for the cosmology dependence of the \citet{Correa_2015} fits. \citet{Wibking_et_al_2019A} report that concentration-mass parameters are highly degenerate with $\delgam$, and therefore marginalizing over the concentration-mass relationship does not degrade cosmological constraints as long as $\delgam$ is included to model uncertainty in the satellite galaxy profile.

In our HOD analysis we choose to consider the galaxy number density $\ngal$ as a parameter because it provides a direct observational constraint on the HOD. Consequently we do not consider $M_{\mathrm{min}}$, $M_{1}$, or $M_{0}$ directly as parameters but instead model the ratios, $M_{1} / M_{\mathrm{min}}$ and $M_{0} / M_{1}$. The actual values of $M_{\mathrm{min}}$, $M_1$, and $M_0$ necessary for implementing our HOD prescription are calculated via numerically integrating over the halo mass function weighted by the galaxy occupation given in equations \ref{eq:cen_HOD} and \ref{eq:sat_HOD}:

\beq
\ngal = \int dM_h \frac{d n_h}{d M_h} \left[ \left< N_{\mathrm{cen}}(M_h) \right> + \left< N_{\mathrm{sat}}(M_h) \right> \right].
\eeq 

In essence we are replacing the central galaxy halo mass scale $M_\mathrm{min}$ with the directly observable number density $\ngal$ in our parameterization. The HOD parameters we consider in the following analysis are therefore $\ngal$, $\siglogM$, $M_1 / M_{\mathrm{min}}$, $M_0 / M_1$, $\alpha$, and $\delgam$. For our fiducial model we adopt values $\ngal = 2.18 \times 10^{-4} \; \invMpchCubed$, $\siglogM = 0.60$, $\satfrac = 0.9$, $\textM0oM1 = -13.7$, $\alpha = 1.60$, and $\delgam = 0.0$. These values are chosen to be consistent with HOD fits to the $M_i < -21.6$ CMASS sample found by \citet{Guo_2014}.

\subsection{Modeling Galaxy Assembly Bias}

Halo assembly bias refers to the phenomenon, observed in simulations, that the clustering of haloes at a fixed mass can depend on properties other than mass \citep[e.g.,][]{Sheth_Tormen_2004, Gao_2005, Harker_et_al_2006,Wechsler_2006, Gao_White_2007, Wang_Mo_Jing_2007, Li_Mo_Gao_2008, Faltenbacher_White_2010, Lacerna_Padilla_2012, Lazeyras_Musso_Schmidt_2017, Villarreal_et_al_2017, Mao_Zentner_Wechsler_2018, Salcedo_2018,  Sato-Polito_et_al_2018, Xu_Zheng_2018}. Denoting such a secondary property as $S_h$ this becomes a simple inequality:

\beq 
\xi_{h}(r|M) \neq \xi_{h}(r|M, S_h). 
\eeq
\emph{Galaxy} assembly bias refers to the potential for the galaxy occupation at a fixed halo mass to depend on other properties $S_h$ that are correlated with clustering \citep[e.g.][]{Zentner_Hearin_vdBosch_2014, Zentner_et_al_2019, Hearin_et_al_2016, Artale_et_al_2018, Zehavi_et_al_2018, Niemiec_et_al_2018, Padilla_et_al_2019, Contreras_et_al_2019}

\beq
\left \langle N_{\mathrm{gal}}(M) \right \rangle \neq \left \langle N_{\mathrm{gal}}(M,S_h) \right \rangle  .
\eeq
These two effects taken in combination will cause a traditional HOD to incorrectly predict the clustering of galaxies \citep[e.g.][]{Zentner_Hearin_vdBosch_2014}. 

To give our model the freedom to account for galaxy assembly bias, we include a parameter $\Qenv$ that allows $M_{\mathrm{min}}$ to vary according to environment \citep{Wibking_et_al_2019A}. Within logarithmic bins of halo mass, we measure matter densities in top-hat spheres of radius $8 \; \Mpch$ and assign them ranks $R_{\delta} \in [0,1]$ according to this environmental density. For each halo the value of $M_{\mathrm{min}}$ is calculated as

\beq
\log M_{\mathrm{min}} = \log M_{\mathrm{min},0} +  \Qenv \left ( R_{\delta} - 0.5 \right ) . \label{eq:Q}
\eeq
Within this prescription, $\Qenv = 0.0$ corresponds to no environmental dependence of galaxy occupation and so is taken as a fiducial value. Although $\emph{halo}$ assembly bias is well predicted from simulations, $\Qenv$ parametrizes $\emph{galaxy}$ assembly bias, which in general will depend on a variety of factors based on the galaxy sample in question. Our prescription is similar to that used by \citet{McEwen_2018}, although we \citep[like][]{Wibking_et_al_2019A} consider the rank in $\delta_8$ rather than the actual value, thus making the dependence of clustering on $\Qenv$ less sensitive to how exactly we measure overdensity.  This prescription is also similar to that of the $\mathcal{A}_{\mathrm{cen}}$ parameter of \citet{Hearin_et_al_2016} and used by \citet{Zentner_et_al_2019}. Although $\mathcal{A}_{\mathrm{cen}}$ is based on halo concentration, both parameters have the effect of boosting the bias on large scales independently from the bias on small scales. The $\Qenv$ prescription makes no specific assumption about what might \emph{cause} galaxy assembly bias and attempts only to describe its effect on clustering.

\subsection{Cluster Modeling}
\label{sec:cluster_model}

We also model clusters for our analysis. The principal challenge in using clusters to constrain cosmology is in accurately characterizing and calibrating the cluster mass-observable relation. Mass calibration is sometimes attempted directly using simulations to predict observables \citep[e.g.][]{Vanderlinde_SPT_2010, Sehgal_ACT_2011}, or by using a sample of clusters with very well measured masses from weak lensing \citep[e.g.][]{Johnston_et_al_2007, Okabe_et_al_2010, Hoekstra_et_al_2015, Battaglia_ACT_2016, vanUitert_et_al_2016, Melchior_DES_2017, Simet_et_al_2017} or X-rays \citep[e.g.][]{Vikhlinin_et_al_2009, Mantz_et_al_2010}. Each of these methods suffers from its own particular limitations. Simulations are limited by our incomplete understanding of baryonic physics, in particular galaxy formation feedback processes. Weak lensing measurements of individual clusters represents a promising method of calibrating the mass-observable relation, but it is limited by signal to noise in addition to systematics such as halo orientation and large scale structure \citep{Becker_Kravstov_2011}. Individual cluster mass measurements using X-rays rely on clusters being in thermal hydrostatic equilibrium and thus can be biased by non-thermal pressure support \citep{Lau_Kravstov_Nagai_2009, Meneghetti_et_al_2010}. 

As discussed in the introduction, our approach here is in some sense a form of weak lensing mass calibration, but instead of calibrating with individual cluster masses we treat the mean tangential shear profile of the full cluster sample as the observable, and the mean and scatter of the mass-observable relation at the selection threshold as parameters to be constrained simultaneously with the cosmological parameters. We study clusters in two redshift bins centred at $z = 0.50$ and $z = 0.30$. In each bin we select clusters via a number density cutoff, $n_{c} = 3.254 \times 10^{-6} \; \invMpchCubed$ at $z = 0.50$ and $n_{c} = 5.846 \times 10^{-6} \; \invMpchCubed$ at $z = 0.30$. For the fiducial cosmology and no mass-observable scatter, these number densities correspond to a to a minimum mass threshold $M_c = 2 \times 10^{14.0} \; h^{-1} \; \Msun$. In an observational sample one cannot select clusters based on mass, only on some observable correlated with mass such as richness, X-ray temperature, X-ray luminosity, or SZ decrement. Having selected clusters above an observable threshold, one can directly measure the space density $n_c$ in a way that is independent of an assumed $\sigma_8$, though there is some dependence on the cosmology assumed to convert redshift and angle separations to comoving distance separations. There will generally be a difference between the space density of clusters in the observed sample and the global average space density of clusters above the observable threshold. We ignore the uncertainty in $n_{c,\mathrm{obs}} - n_{c, \mathrm{global}}$ in our analysis, but we show in \cref{subsec:nc_uncertainties} that it should have negligible impact.

We characterize the cluster-mass observable relation as a linear relation with a constant lognormal scatter $\siglnMc$,

\beq 
\ln M_{\mathrm{obs}} = \ln M_c + \siglnMc X,
\eeq
where $X \sim \mathcal{N}(0,1)$. Other studies \citep[e.g.][]{Murata_et_al_2018} have chosen more complicated functional forms to characterize this relation and have allowed the scatter to vary with mass. For our purposes this simple form suffices because we care only about scatter of clusters across the single selection boundary. Analyses of counts in multiple bins may in principle use more information, but they also require more nuisance parameters to describe the mass-observable relation (see \cref{subsec:Cdef_Bin} below).

\begin{figure}
\centering
\includegraphics[width=0.45\textwidth]{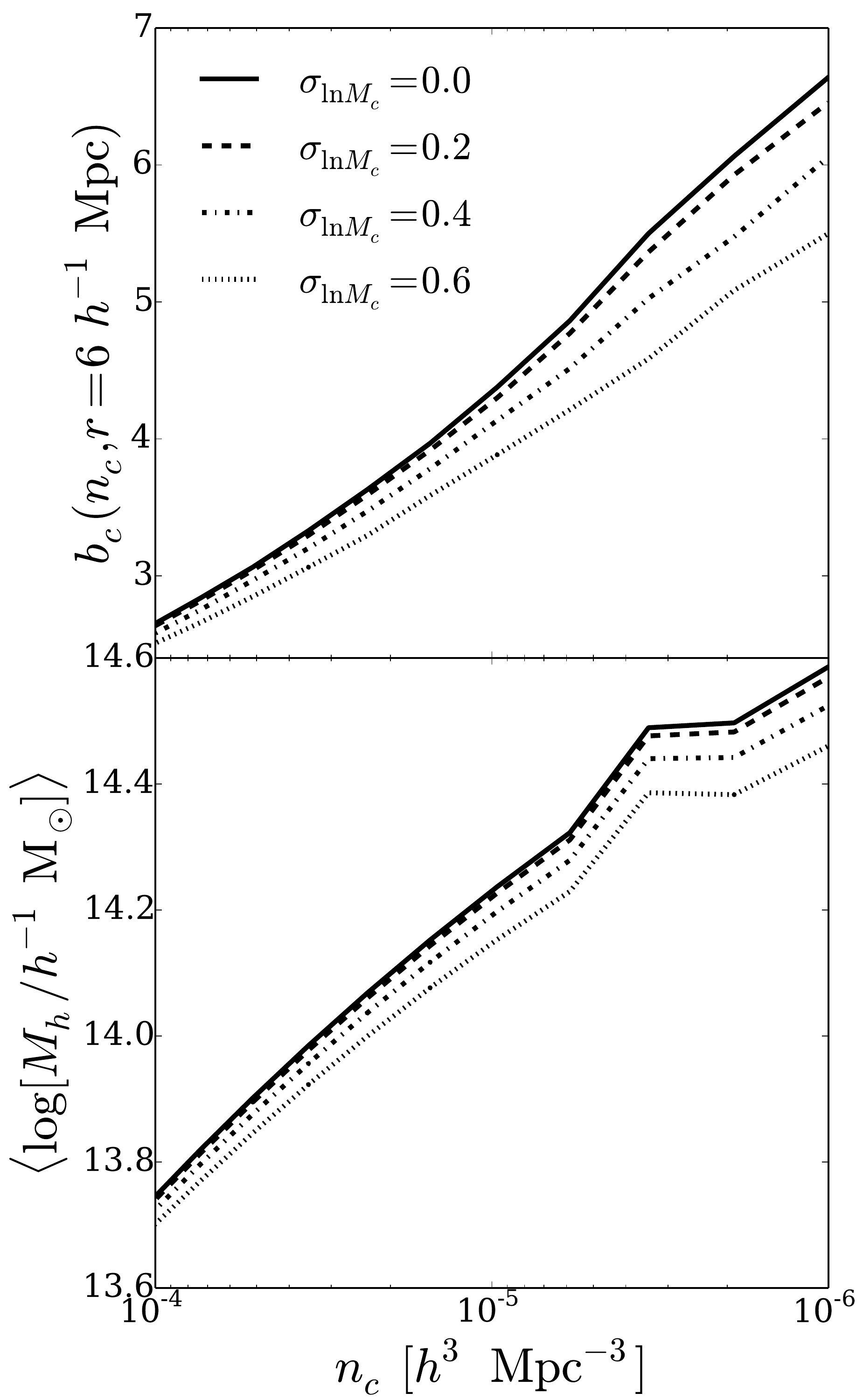}
\caption{The effect of $\siglnMc$ on the bias and mean mass of haloes as a function of number density threshold. The case of $\siglnMc = 0.0$ shows the expected behavior of the bias; an increasing function of mass that is shallow at small masses but becomes increasingly steep for increasing mass. With large $\siglnMc$, the average bias of a sample with the same number density (top axis) decreases.}
\label{fig:scatter_bias}
\end{figure} 

The scatter is a critical nuisance parameter because it is largely degenerate with $\Omega_m$ and $\sigma_8$. Because lower mass haloes are more numerous, scatter tends to replace haloes above the sample mass threshold with haloes of a lower mass and lower clustering bias. In a sample of a given $n_c$, a higher $\siglnMc$ leads to lower mean mass and lower clustering. This effect is shown in Figure \ref{fig:scatter_bias}, where the bias is calculated by averaging the cluster bias, 

\beq
b_c = \sqrt{\frac{\xi_c(r)}{\xi_{mm}(r)}},
\eeq
over the 20 realizations of the fiducial cosmology.

\section{Derivatives of Observables with Respect to Parameters}
\label{sec:derivs}

\subsection{Clustering and Weak-Lensing Statistics}
\label{subsec:clustering_stats}

\begin{figure*}
\centering
\includegraphics[width=1.0\textwidth]{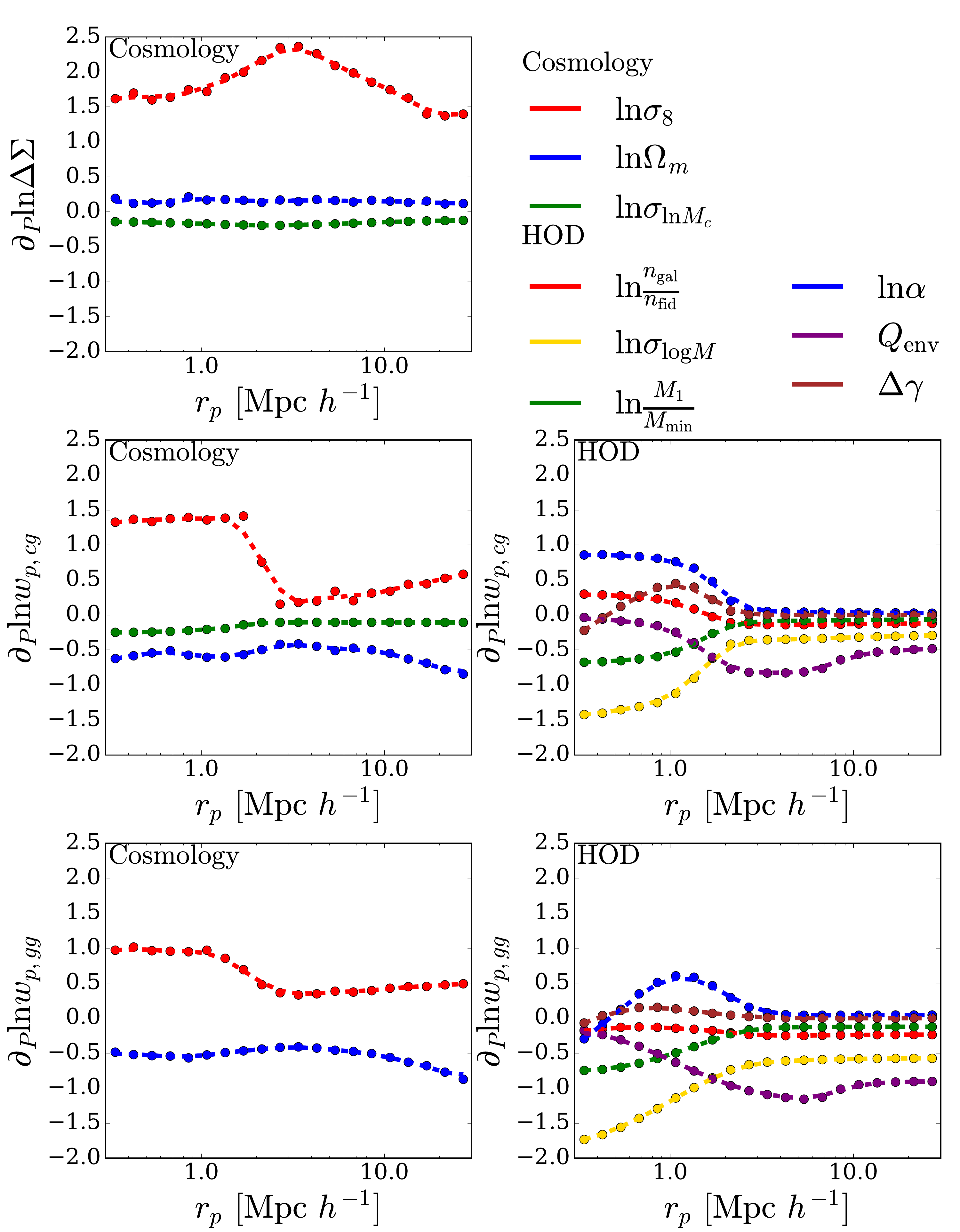}
\caption{Logarithmic derivatives of the clustering observables $w_{p,cg}$ and $w_{p,gg}$ with respect to HOD parameters (right panels) and cosmological parameters (left panels) and of the lensing observable $\Delta \Sigma$ with respect to cosmological parameters (top left). Points represent values calculated from the simulations, and the dashed lines show the result of smoothing the derivatives as explained in \cref{subsec:clustering_stats}.}
\label{fig:derivatives}
\end{figure*}

The amplitude of spatial clustering can be measured by correlation functions. In particular we will use two point cross-correlation functions to study the amplitude of cluster-galaxy and cluster-matter clustering. The two point cross-correlation function, $\xi_{AB}(r)$, is defined by the joint probability of finding objects in two volume elements ($\delta V_A$,$\delta V_B$) separated by some distance $r$,

\beq
\delta P = n_A n_B \delta V_A \delta V_B \left[ 1 + \xi_{AB}(r) \right],
\eeq
where $n_A$ and $n_B$ are the respective number densities of the sets of objects considered \citep{Peebles_tLSSotU}. Written this way it is clear that the correlation function measures an excess in spatial clustering from that of a random distribution of points. In practice we estimate the cross-correlation function using the Landy-Szalay estimator \citep{Landy-Szalay_1993},

\beq
\xi_{AB}(r) = \frac{AB(r) - AR(r) - BR(r) + RR(r)}{RR(r)},
\eeq
where $AB(r)$ is the observed number of A-B pairs with separation, $r$, $AR(r)$ and $BR(r)$ are the number of A-random and B-random pairs respectively, and $RR(r)$ is the expected number of such pairs in a random sample with the same respective number densities and volume geometry. When the volume being considered is periodic, we analytically calculate the expected number of random pairs as $RR(r) = AR(r) = BR(r) = 4 \pi n_A n_B r^2 dr$. In such a case the Landy-Szalay estimator is equivalent to the ``natural'' estimator, 

\beq
\xi_{AB}(r) =  \frac{AB(r)}{RR(r)} - 1.
\eeq

We use {\sc{corrfunc}} \citep{Sinha_2017} to compute the real space cluster-galaxy crosscorrelation function $\xi_{cg}(r)$ and cluster-matter crosscorrelation function $\xi_{cm}(r)$, in 50 equal logarithmically spaced bins covering scales $0.05 < r < 125.0 \; \Mpch$, averaging over 20 HOD realizations at each point in parameter space. With these real-space correlation functions we calculate the more observationally-motivated quantities $w_{p,cg}(r_p)$, $w_{p,gg}(r_p)$ and $\Delta \Sigma (r_p)$. Neglecting sky curvature, residual redshift-space distortions, and higher-order lensing corrections, these can be calculated by integrating over the appropriate correlation functions:

\begin{align}
\begin{aligned} 
w_{p,AB}(r_p) &= 2 \int_0^{\pimax} \xi_{AB} \left(r, \pi \right) d \pi,
\end{aligned}
\end{align}

\begin{align}
\begin{aligned}
\Delta \Sigma (r_p) &= \Omega_m \rho_{\mathrm{crit}}  \left[ \frac{2}{r_p^2} \int_{0}^{r_p} r' w_{p,cm}(r') dr' - w_{p,cm}(r_p) \right].
\end{aligned}
\label{eq:delsigma}
\end{align}
For a specified distribution of source redshifts (i.e., lensed galaxies), the observable tangential shear profile is simply proportional to $\Delta \Sigma(r_p)$. Uncertainty in the source redshift distribution leads to uncertainty in $\Delta \Sigma(r_p)$, but we do not consider this survey-specific problem here.

To avoid the effect of redshift distortions on clustering measurements one would ideally want $\pimax \rightarrow \infty$. However a finite $\pimax$ can be sufficient to measure $w_p$ to the required precision depending on survey properties. In all that follows we adopt $\pimax = 100.0 \; \Mpch$. In DES Science Verification Data, redMaGiC selected galaxies within the range $0.2 < z < 0.8$ have errors on photmetric redshifts $1 + z_p$ on the order of $1-2 \%$ \citep{Rozo_DES_2016}. In the redshift range we consider, $0.35 < z < 0.55$, errors of this magnitude correspond to errors in $\pi$ on the order of $30.0 \; \Mpch$, well below our value of $\pimax$.

We calculate partial derivatives of observables with respect to model parameters, using finite differences centred at the fiducial parameter values with step sizes in cosmology determined by our simulation grid and step sizes in HOD motivated by the fit errors of \citet{Guo_2014}. For each of the parameters these steps (while holding all else equal) are $\ngal/\nfid = 1.0 \pm 0.1$, $\siglogM = 0.60 \pm 0.05$, $\Qenv = 0.0 \pm 0.1$, $\satfrac = 0.9 \pm 0.1$, $\alpha = 1.6 \pm 0.1$, $\delgam = 0.0 \pm 0.2$, $\siglnMc = 0.4 \pm 0.2$, $\Omega_m = 0.314 \pm^{0.30}_{0.26}$, and $\sigma_8 = 0.83 \pm 0.05$.  When forecasting in subsequent sections we additionally smooth these measured derivatives as a function of $r_p$ with a Savitsky-Golay filter. 
 
 Figure \ref{fig:derivatives} shows the result of our direct calculation of derivatives. The right column of panels shows derivatives of $w_{p,cg}$ and $w_{p,gg}$ with respect to HOD parameters. We observe that within the 1-halo regime there is a great deal of scale dependent behavior. The left column shows cosmological parameter derivatives for $w_{p,cg}$, $w_{p,gg}$ and $\Delta \Sigma$. We group $\siglnMc$ with the cosmological parameters. Recall that all derivatives are evaluated at fixed cluster number-density (not fixed mass threshold) and that the $\Omega_m$ derivative is evaluated at fixed $\Omega_m h^2$. It is the cosmological parameter derivatives, which we cannot average over 20 realizations, that most require smoothing, which conservatively removes noise-like features that could artificially improve our parameter forecasts.

\subsection{ Effect of Parameter Variations}
\label{sec:param_var}

\begin{figure*}
\centering
\includegraphics[width=1.0\textwidth]{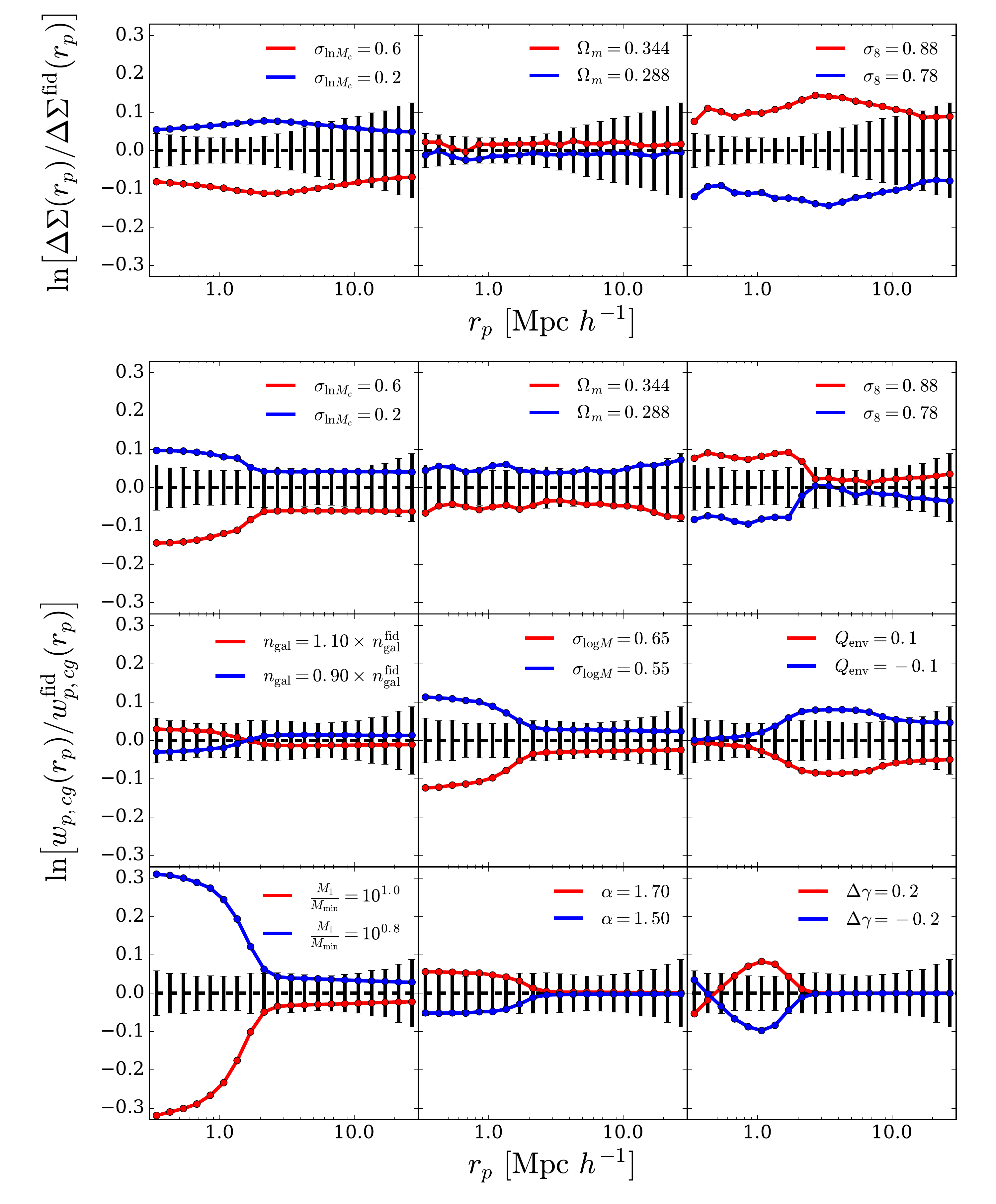}
\caption{Fractional changes to the lensing observable $\Delta \Sigma$ (top row) and the cluster galaxy cross correlation function $w_{p,cg}$ (bottom three rows) induced by changes in cosmological parameters $\Omega_m$ or $\sigma_8$, the mass observable scatter $\siglnMc$, or the six galaxy HOD parameters. In each panel red and blue curves show the predicted change of the observable for the parameter values indicated in the panel legend which are perturbed symmetrically about our fiducial parameter choice. Error bars show the (square root of the) diagonal elements of the covariance matrix estimated for a DES-like cluster and weak lensing survey (\cref{sec:cov}). Curves for $\Omega_m$ and $\sigma_8$ are noisier because they use a single realization of the initial Fourier phases, while HOD and $\siglnMc$ derivatives are averaged over 20 realizations of the fiducial cosmology. Changes to $\Omega_m$ are made at fixed $\Omega_m h^2$.}
\label{fig:cg_frac_diffs}
\end{figure*}

\begin{figure*}
\centering
\includegraphics[width=1.0\textwidth]{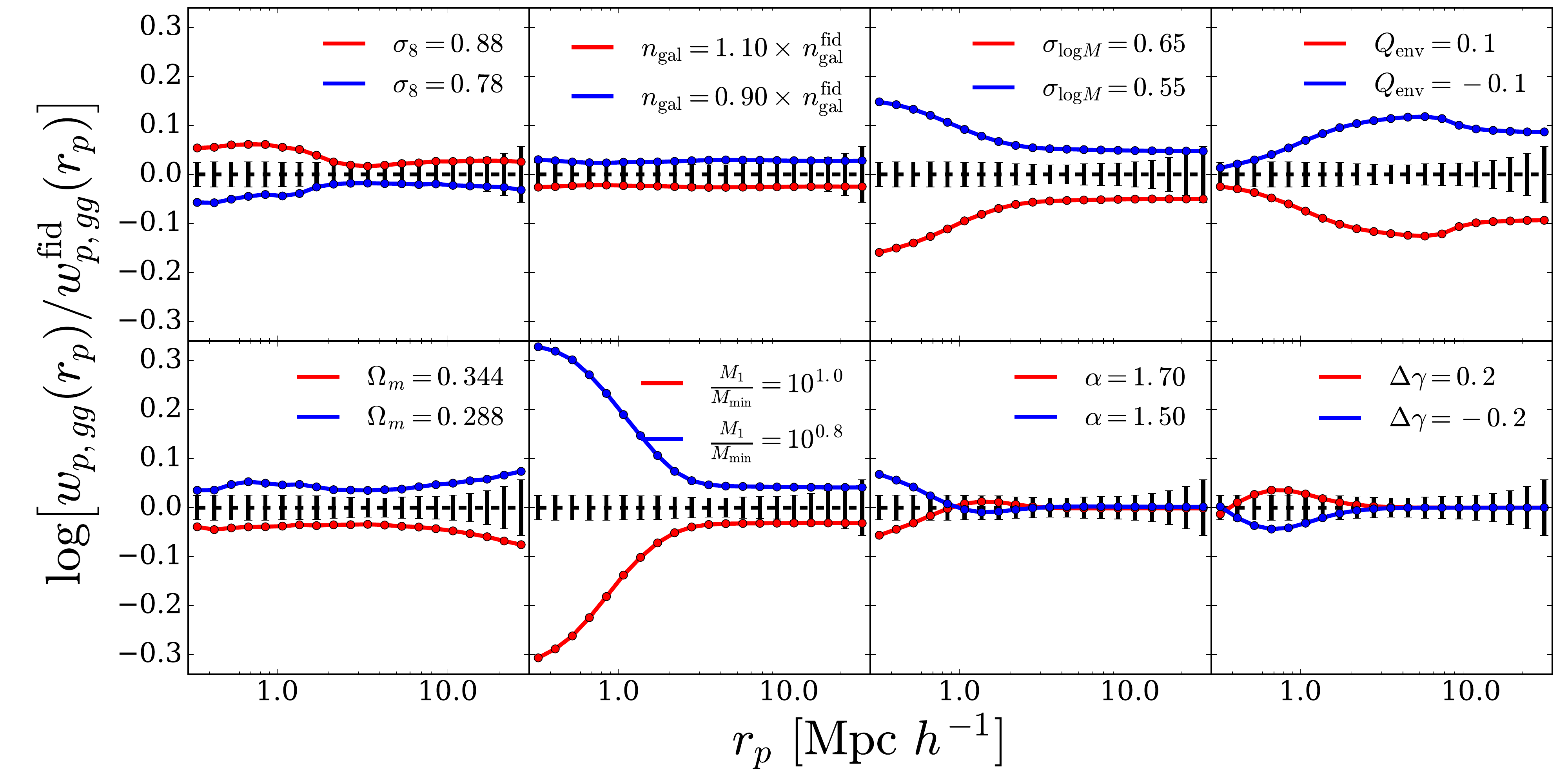}
\caption{The logarithm of the ratios of the projected galaxy-galaxy correlation function, $w_{p,gg}(r_p)$ for changes in our extended HOD and cosmological parameters. Error bars represent the fractional standard deviation from the elements of our model covariance matrix.}
\label{fig:gg_frac_diffs}
\end{figure*}

Instead of discussing the derivatives directly, we examine the impact of specified parameter choices on $\Delta \Sigma$ (Figure \ref{fig:cg_frac_diffs}, top), $w_{p,cg}$ (Figure \ref{fig:cg_frac_diffs}, bottom), and $w_{p,gg}$ (Figure \ref{fig:gg_frac_diffs}). Note that HOD parameters have no impact on $\Delta \Sigma$ and that $\siglnMc$ has no impact on $w_{p,gg}$. In each panel, red (blue) curves show the effect of increasing (decreasing) the indicated parameter relative to the fiducial value. For comparison, error bars show the diagonal elements of the covariance matrix estimated in \cref{sec:cov}, motivated roughly by a DES-like cluster and weak lensing survey. 

Beginning with $\Delta \Sigma$, we see that increasing either $\Omega_m$ or $\sigma_8$ increases the predicted $\Delta \Sigma (r_p)$ at all scales. This trade off produces the well known $\sigma_8 - \Omega_m$ degeneracy, but the detailed shape of this degeneracy depends on what one holds fixed when changing $\Omega_m$. We have chosen to fix $\Omega_m h^2$, which is well constrained by the CMB, so the shape of the power spectrum in $\Mpch$ becomes ``bluer'' as $\Omega_m$ increases (i.e., more small scale power power relative to the normalization at $8 \; \Mpch$). With this choice, the impact of a $19\%$ change in $\Omega_m$ (red vs. blue curves in the $\Omega_m$ panel) is much smaller than the impact of a $13\%$ change (red vs. blue) in $\sigma_8$. The impact of a $\sigma_8$ change is moderately scale-dependent, with the largest change to $\Delta \Sigma$ at scales of a few $\Mpch$. Figure \ref{fig:cg_frac_diffs} clearly illustrates the degeneracy between $\sigma_8$ and $\siglnMc$, with $\siglnMc$ depressing $\Delta \Sigma$ on all scales by reducing the average mass and clustering bias of clusters above the selection threshold (see Figure \ref{fig:scatter_bias}). Constraining $\sigma_8$ with cluster weak lensing requires external constraints on $\siglnMc$, which in our analysis will come from $w_{p,cg}$ and $w_{p,gg}$.

For fixed HOD parameters, the impact of $\siglnMc$ on $w_{p,cg}$ is qualitatively similar to that of $\Delta \Sigma$, but the scale-dependence is different, with a more prominent boost in the 1-halo regime for reduced $\siglnMc$. The impact of $\Omega_m$ or $\sigma_8$ changes is affected by our decision to treat $\ngal$ rather than $M_{\mathrm{min}}$ as the fixed HOD parameter (though we allow it to vary in our multi-parameter fits in \cref{sec:forecasts}). Boosting $\Omega_m$ or $\sigma_8$ shifts the halo mass function upward in the mass regime relevant for CMASS-like galaxies. As a result, $M_{\mathrm{min}}$ shifts upwards to keep $\ngal$ fixed, but the bias factor of these halos may still be reduced if $M_{\mathrm{min}}/ M_{\mathrm{nl}}$ is lower, where $M_{\mathrm{nl}}$ is the non-linear mass scale defined by $\sigma_8(M_\mathrm{nl}) \approx 1$. Similarly because we hold the cluster space density $\ncluster$ fixed, the cluster mass threshold drops relative to $M_{\mathrm{nl}}$ when $\sigma_8$ is increased. 

For perturbations about our fiducial model, increasing $\Omega_m$ depresses $w_{p,cg}$ at all scales; the sign of this effect is opposite to that of $\Delta \Sigma$ because the excess surface density is proportional to $\Omega_m$ (eq. \ref{eq:delsigma}) and independent of galaxy bias. Increasing $\sigma_8$ boosts the number of high-occupancy haloes and therefore boosts $w_{p,cg}$ in the 1-halo regime, but on large scales the increase of $\xi_{mm}$ is nearly cancelled by the reduction in galaxy and cluster bias. In detail, at a scale of $10 \; \Mpch$, raising $\Omega_m$ from 0.314 to 0.344 changes ($\xi_{mm}$, $b_c$, $b_g$) by ($-3.35\%$, $-1.72\%$, $+1.83\%$). Raising $\sigma_8$ from 0.83 to 0.88 changes ($\xi_{mm}$, $b_c$, $b_g$) by ($+12.09\%$, $+10.35\%$, $-10.50\%$).

The third row of Figure \ref{fig:cg_frac_diffs} shows the impact of parameters that directly affect the central galaxy occupation, with cosmological parameters and $\siglnMc$ now fixed to their fiducial values. Raising $\ngal/n_{\mathrm{fid}}$ leads to a reduction of $M_{\mathrm{min}}$, causing a drop in the large scale galaxy bias that reduces $w_{p,cg}$. Because we keep $M_1 / M_{\mathrm{min}}$ fixed, the number of satellites in massive halos goes up, boosting $w_{p,cg}$ in the 1-halo regime. Increasing $\siglogM$ allows more halos with $M < M_\mathrm{min}$ to host central  galaxies. The value of $M_\mathrm{min}$ must be raised to keep $\ngal$ fixed, but the average galaxy bias still decreases because of the larger number of centrals hosted by lower mass haloes. Because $M_1 / M_\mathrm{min}$ is fixed, the number of satellites in massive haloes declines, and the 1-halo regime of $w_{p,cg}$ is depressed much more strongly than the large scale regime.

A positive value of our environmental dependence parameter $\Qenv$ raises $M_\mathrm{min}$ in high density regions (eq. \ref{eq:Q}). It therefore reduces galaxy numbers in overdense regions (and vice versa), so it reduces the galaxy bias and depresses $w_{p,cg}$ on large scales. A negative value of $\Qenv$ boosts large scale clustering. In the 1-halo regime, galaxy clustering depends on integrals of $P(N|M)$ over the halo mass function \citep[e.g.][]{Berlind_2002}, without reference to the halo environment. We therefore expect the impact of galaxy assembly bias on galaxy clustering to decline on small scales, as seen in Figures \ref{fig:cg_frac_diffs} and \ref{fig:gg_frac_diffs}. However, the particular form of scale dependence doubtless depends to some degree on our choice of $8 \; \Mpch$ as the scale for defining environment. The addition of $\Qenv$ to the HOD parameter set allows the large scale galaxy bias to decouple from the small and intermediate scale clustering constraints on other HOD parameters. Further work will be needed to see if this added freedom is sufficient to capture the impact of all realistic scenarios for  galaxy assembly bias. 

The bottom row of Figure \ref{fig:cg_frac_diffs} shows the impact of HOD parameters that directly impact the satellite populations, though there is a weak link to central galaxies through $\ngal$. Raising $M_1/M_\mathrm{min}$ reduces the occupancy of high mass haloes and strongly depresses $w_{p,cg}$ in the 1-halo regime. There is a weak boost on large scales coming from the contribution of satellites to $b_g$. Increasing $\alpha$, and thus boosting the occupancy of the highest mass haloes relative to haloes with $M \sim M_1$, has negligible impact on large scales and only a small impact (for $\Delta \alpha$ = 0.1) in the 1-halo regime. A positive $\delgam$ preferentially moves satellites to larger $r/R_\mathrm{vir}$ (eq. \ref{eq:delgam}), effectively decreasing halo concentration. The number of satellites per halo does not change, so the boost of $w_{p,cg}$ at $r_p \sim 1 \; \Mpch$ is compensated by a reduction at the smallest scales.

The impacts of parameters on $w_{p,gg}$ (Figure \ref{fig:gg_frac_diffs}) are qualitatively similar to the impacts on $w_{p,cg}$. The dependence on $\ngal$ is different, as an increase of $\ngal$ depresses $w_{p,gg}$ on all scales, without the crossover in the 1-halo regime see for $w_{p,cg}$. In principle this difference could be useful in constraining $\ngal$, but because it is a directly measured quantity itself, an indirect constraint from clustering will probably not reduce its uncertainty. The impact of $\alpha$ is also opposite in $w_{p,gg}$ and $w_{p,cg}$. Because of our constant $\ngal$ constraint, increasing $\alpha$ actually decreases $w_{p,gg}$ on small scales because $\Mmin$ and $M_{1}$ increase and the number of satellites declines. However, $w_{p,cg}$ weights the highest mass haloes more strongly so higher $\alpha$ increases $w_{p,cg}$ at small scales. The impact of $\Qenv$ is somewhat stronger for $w_{p,gg}$ than for $w_{p,cg}$, probably because the first is proportional to $b_g^2$ and the second to $b_g$.

From Figures \ref{fig:cg_frac_diffs} and \ref{fig:gg_frac_diffs} we can see how the addition of $w_{p,cg}$ and $w_{p,gg}$ can improve the cosmological constraints from cluster weak lensing. With $\Delta \Sigma$ measurements alone, deriving constraints on $\sigma_8$ and $\Omega_m$ requires a tight external prior on $\siglnMc$, since the impact of mass-observable scatter is largely degenerate with the impact of $\sigma_8$. Measurements of $w_{p,cg}$ provide an independent constraint on $\siglnMc$, but the impact of $\siglnMc$ is degenerate with that of some HOD parameters, especially $\siglogM$ and $\satfrac$, which have qualitatively similar scale dependence. Measurements of $w_{p,gg}$ provide constraints on these HOD parameters that are independent of $\siglnMc$. Therefore the two galaxy clustering measures together constrain $\siglnMc$, which allows $\Delta \Sigma$ to constrain cosmological parameters. Our forecasts in \cref{sec:forecasts} bear out this interpretation. In particular, we find that cosmological constraints from the combination of $\Delta \Sigma$, $w_{p,cg}$, and $w_{p,gg}$ are much stronger than those from any two of these statistics alone.

\section{Estimating the Measurement Covariance Matrix}
\label{sec:cov}

To forecast cosmological parameter constraints, or to derive constraints from observational data, we require derivatives of observables with respect to parameters and an error covariance matrix for the observables themselves. Having adressed the former in \cref{sec:derivs}, we now turn to the latter. As cosmological surveys increase in size and precision, the challenge of constructing accurate covariance matrices grows more severe. One general approach is to make many realizations of a simulated data set, but for large surveys this may be computationally infeasible. Analytic approximations avoid computational limits and the noise and bias that can arise from a small number of realizations, but they may be inaccurate in a regime where non-Gaussianity of the matter or galaxy fields is important. For a given observational data set or mock data set, one can also use subsampling or jackknife methods to estimate statistical errors and their covariances. 

For this paper we use a combination of numerical and analytic methods to estimate the covariance matrix. For the $w_{p,cg}$ and $w_{p,gg}$ statistics, we compute diagonal elements of the covariance matrix using subsamples of our 20 {\sc{abacus}} simulations of the fiducial cosmology, and we compute off-diagonal elements from an analytic correlation matrix based on the methods of \citet{Krause_Eifler_2017}. For $\Delta \Sigma$ the problem is more challenging, because at large $r_p$ even the diagonal errors are dominated by noise in the cosmic shear from structure over the entire redshift range of the lensing survey, which we cannot estimate from volumes the size of our {\sc{abacus}} boxes. We give a full discussion of our calculation of cluster weak lensing covariance matrices in a separate paper \citep{Wu_et_al_2019}. In brief, we use the analytic formalism of \citet[][]{Jeong_et_al_2009}, similar to that of \citet{Marian_Smith_Angulo_2015} and \citet{Singh_et_al_2017}, to compute the covariance at large scales, and we use the {\sc{abacus}} simulations to compute the covariance at small scales, merging them in a consistent way and adding shape noise as a separate component. We show that this approach reproduces covariances measured from the weak lensing  survey simulations of \cite{Takahashi_et_al_2017}, based on ray tracing through a matter field constructed by replicating N-body simulations.

Our fiducial forecast is motivated, somewhat loosely, by the properties of the DES cluster and weak lensing survey and the BOSS CMASS galaxy redshift survey. Our assumed parameters are summarized in \cref{table:covariance} We consider two redshift bins for the clusters, $z = 0.15-0.35$ and $z = 0.35-0.55$, and a survey area of $\Omega = 5000 \; \mathrm{deg}^2$. The comoving survey volumes are $V_s = 4.071 \times 10^8 \; \MpchCubed$ and $V_s = 1.042 \times 10^9 \; \MpchCubed$, respectively. We model these two bins using the $z = 0.3$ and $z = 0.5$ outputs of the {\sc{abacus}} simulations, ignoring the effects of evolution across the redshift bin. The mean redshifts of DES redMaPPer selected clusters \citep[][]{McClintock_DES_2019} in these ranges are 0.25 and 0.44, slightly lower than our simulation outputs. As previously noted, the space density of clusters for our adopted threshold is $n_c = 5.846\times10^{-6}\; \MpchCubed$ at $z =  0.3$ and $3.254\times10^{-6} \; \MpchCubed$ at $z = 0.5$, making the total cluster numbers in the model survey $n_c V_s = 7781$ and $4331$, respectively.

Based on the source redshift distribution from \citet{Rozo_2011} we compute mean redshifts $\left< z_s\right> = 0.89$ and $\left< z_s \right> = 0.99$ for sources lensed by the two cluster samples, and source surface densities $\Sigsrc = 9.0 \; \mathrm{arcmin}^{-2}$ and $7.2 \; \mathrm{arcmin}^{-2}$, respectively. For simplicity, we compute the covariance matrix for each cluster sample assuming all source are at the mean redshift, i.e., using a single value of $\Sigcrit$. We assume a shape noise per galaxy of $\sigma_\gamma = 0.3$.

\subsection{Analytic estimation}
\label{sec:analytic}

Our discussion here is closely modeled on that of \citet[][also see: \citet{Cooray_Hu_2001,Marian_Smith_Angulo_2015, Krause_Eifler_2017}]{Singh_et_al_2017} Following the arguments of \citet{Krause_Eifler_2017}, we have the Fourier-space Gaussian covariance between two power spectra in a single redshift bin,

\begin{align}
    \mathrm{cov}&\left(P_{AB}(k_m), P_{CD}(k_n) \right) = \frac{ (2 \pi)^3 \delta(k_m - k_n)}{ V_s \left( 4 \pi k_m^2 \right)} \nonumber \\
    &\times [  \left(P_{AC}(k_m) + \delta_{AC} N_A \right)\left(P_{BD}(k_n) + \delta_{BD} N_B \right) \nonumber \\
    &+ \left(P_{AD} (k_m) + \delta_{AD} N_A \right)\left( P_{BC}(k_n) + \delta_{BC} N_B \right)],
\end{align}
where $V_s$ is the survey volume for which we are calculating the measurement covariance. If we desire the covariance for projected correlation functions we must integrate over the line of sight window functions and convert from Fourier-space to configuration space. This is accomplished by multiplying the Fourier space covariance by the Fourier transforms of circles (zero-order Bessel functions of the first kind\footnote{Recall: $J_n(x) = \int_{-\pi}^{\pi} \frac{d \phi}{2 \pi} e^{i \left[x \mathrm{sin}(\phi) - n \phi \right]}$.}) with radii $r_i$ and $r_j$ and integrating over all modes:

\begin{align}
    \mathrm{cov}(&w_{p,AB}(r_{p,i}), w_{p,CD}(r_{p,j}) ) = \frac{2 \pimax}{V_s} \int_0^{\infty} \frac{k dk}{2 \pi} J_0(k r_i) J_0(k r_j) \nonumber \\
    &\times [  \left(P_{AC}(k) + \delta_{AC} N_A \right)\left(P_{BD}(k) + \delta_{BD} N_B \right) \nonumber \\
    &+ \left(P_{AD} (k) + \delta_{AD} N_A \right)\left( P_{BC}(k) + \delta_{BC} N_B \right)],
    \label{eq:gen_cov}
\end{align}
where we have assumed top hats for the line of sight window functions, $2 \pimax = \int_{-\pimax}^{\pimax} d \pi W_{AB}(\pi) W_{CD} (\pi)$, as is the case for projected correlation functions. To obtain the covariance in bins we simply replace the Bessel functions in the above expression by the corresponding bin averaged Bessel functions

\beq
    \hat{J}_0 (\rmin, \rmax, k) = \frac{2 \left[ \rmax J_1 (k \rmax) - \rmin J_1 (k \rmin) \right]}{k \left( \rmax^2 - \rmin^2 \right)},
\eeq
where $\rmin$ and $\rmax$ are the inner and outer boundaries of a bin for which the covariance is being measured. Applying these expressions we can write the covariance for $w_{p,cg}$ and $w_{p,gg}$ as well as the cross observable covariance:

\begin{align}
\mathrm{cov} \left( w_{p,gg}(r_{p,i}), w_{p,gg}(r_{p,j}) \right) &= \frac{ 4 \pimax}{V_s} \int_0^\infty \frac{k dk}{2 \pi} \hat{J}_0 (k r_i) \hat{J}_0 (k r_j) \\
&\times \left[  P_{gg}(k) + \frac{1}{n_g} \right]^2, \nonumber
\end{align}

\begin{align}
\mathrm{cov} ( &w_{p,cg} (r_{p,i}), w_{p,cg} (r_{p,j}) ) = \frac{2 \pimax}{V_s} \int_0^{\infty} \frac{k \, dk}{2\pi} \, \hat{J}_0(k r_i) \hat{J}_0(k r_j) \\
&\times \left[ \left(P_{cc}(k) + \frac{1}{n_c} \right) \left( P_{gg}(k) + \frac{1}{n_g} \right) + P^2_{cg}(k) \right], \nonumber
\end{align}

\begin{align}
\mathrm{cov} \left( w_{p,cg} \left( r_{p,i} \right), w_{p,gg} \left(r_{p,j}\right) \right) = \frac{4\pimax}{V_s} \int_0^{\infty} \frac{k dk}{2\pi}  \hat{J}_0(k r_i)  \hat{J}_0(k r_j) \\
\times \left[ P_{cg}(k)  \left( P_{gg}(k) + \frac{1}{n_g} \right) \right]. \nonumber
\end{align}

\begin{table*}
   \centering
   \begin{tabular}{ccc p{3.2cm}}
      \hline
      Quantity & Bin 1 & Bin 2 & Description \\
 	\hline
      $\ngal$ &  $2.18 \times 10^{-4} \; \invMpchCubed$ & $2.18 \times 10^{-4} \; \invMpchCubed$ & galaxy number density \\
      $\ncluster$ & $5.846 \times 10^{-6} \; \invMpchCubed$ &  $3.254 \times 10^{-6} \; \invMpchCubed$ & cluster number density \\
      $\pimax$ & $100.0 \; \Mpch$ & $100.0 \; \Mpch$ & max. projection length \\
      $\Omega$ & $5000 \; \mathrm{deg}^2 $ & $5000 \; \mathrm{deg}^2 $ & survey area \\
      $\left[z_{\mathrm{min}}, z_{\mathrm{max}} \right]$ & $[0.15, 0.35]$ &  $[0.35, 0.55]$ & survey redshift limits \\
      $V_s$ &  $4.071 \times 10^{8} \; \MpchCubed$ & $1.042 \times 10^{9} \; \MpchCubed$  & survey volume \\
      $\sigma_\gamma$ & 0.3 & 0.3 & shape noise per galaxy \\
      $\Sigsrc$ & $ 9.0 \; \mathrm{arcmin}^{-2}$ &  $ 7.2 \; \mathrm{arcmin}^{-2}$ & source density \\
      $\left<z_L\right>$ & 0.3 & 0.5 & mean lens redshift \\
      $\left<z_S\right>$ & 0.89 & 0.99 & mean source redshift \\
	\hline
   \end{tabular}
   \caption{Values used in covariance estimation, ordered by mention in text. These parameters are chosen to approximate a DES-like cluster and weak lensing survey with a CMASS-like galaxy sample.}
\label{table:covariance}
\end{table*}

\begin{figure}
\centering
\includegraphics[width=0.45\textwidth]{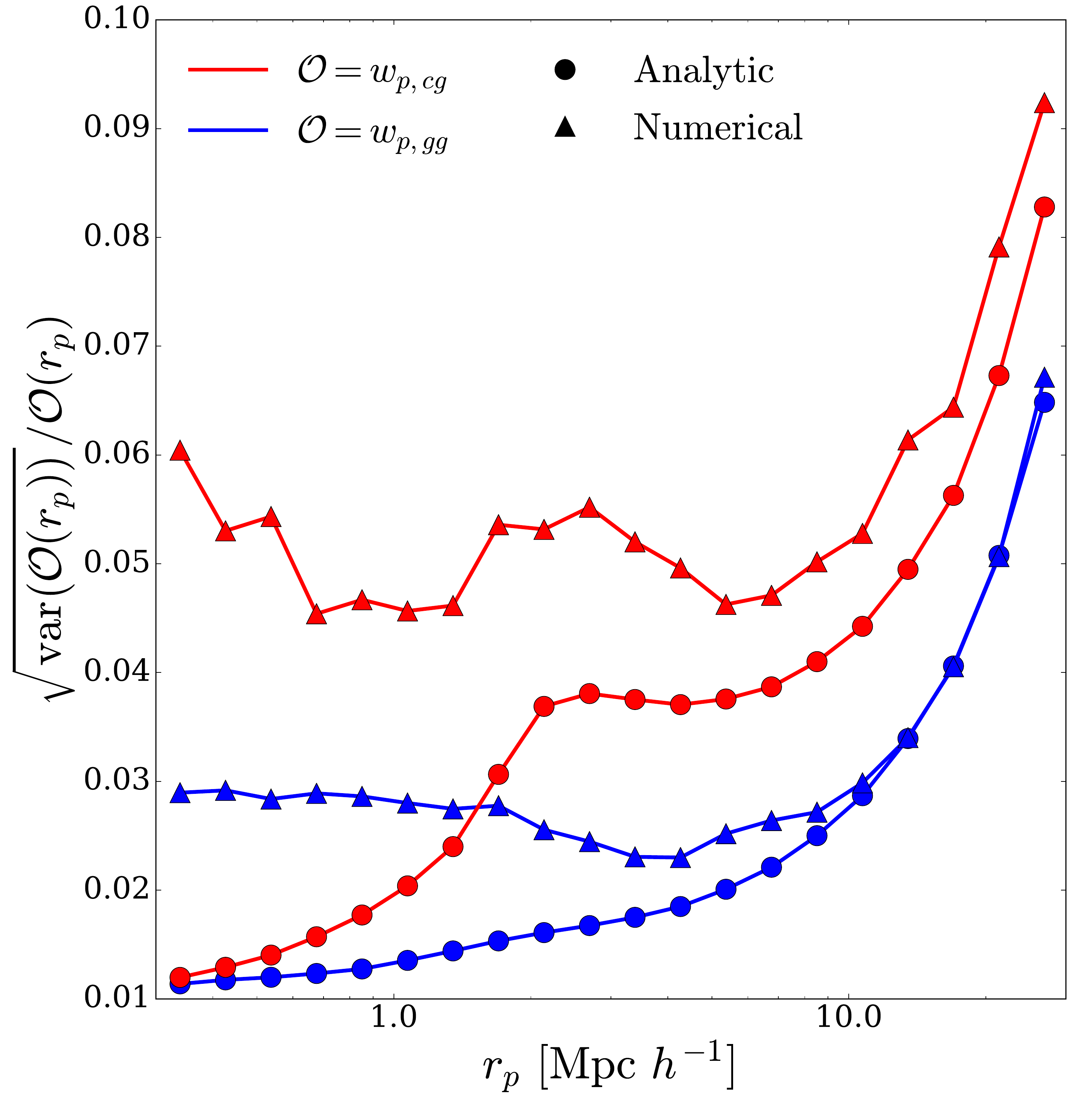}
\caption{Diagonal elements of the covariance matrix, converted to fractional errors, for the observables $w_{p,cg}$ (red) and $w_{p,gg}$ (blue). Circles and triangles show analytic and numerical estimates, respectively, for the fiducial survey parameters in the $z = 0.35-0.55$ bin.}
\label{fig:diag_comp}
\end{figure}

The analytic formalism for $\Delta \Sigma$ covariances is similar, though second-order Bessel functions replace zeroth-order because of the bilateral symmetery of galaxy shears, and shape noise $\sigma_\gamma^2/\Sigsrc$ plays the role of galaxy shot noise $1/n_g$. The formalism is also more complicated because the lensing redshift kernel is inherently broad, so one cannot consider power spectra at a single redshift and $\pimax$ much smaller than the survey depth. We leave further discussion of the analytic $\Delta \Sigma$ covariance and our method of merging it with the numerical covariance matrix to our companion paper \citet{Wu_et_al_2019}.

We calculate all of these contributions to the measurement covariance in 20 logarithmically spaced bins in the range $r_p = 0.3 - 30.0 \; \Mpch $ with $\pimax = 100 \; \Mpch$, using non-linear power spectra calculated from our simulations.  
 
\begin{figure*}
\centering
\includegraphics[width=1.0\textwidth]{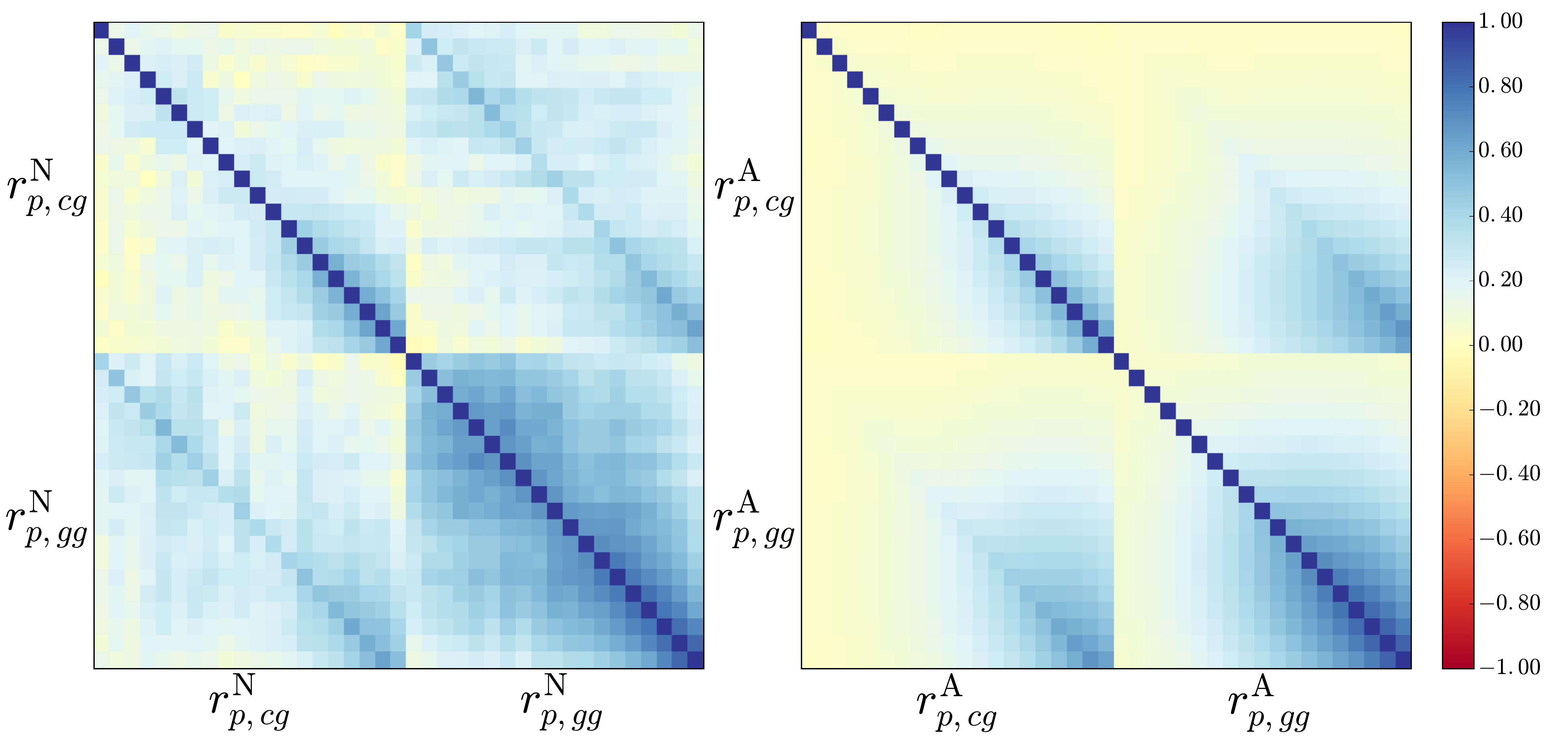}
\caption{Correlation matrix of the clustering observables in our data vector calculated numerically (left) and analytically (right). While there is qualitative agreement of the two estimates, the numerical estimate shows somewhat stronger off-diagonal structure and is noisier.}
\label{fig:corr_matrix}
\end{figure*}

\begin{figure*}
\centering
\includegraphics[width=1.00\textwidth]{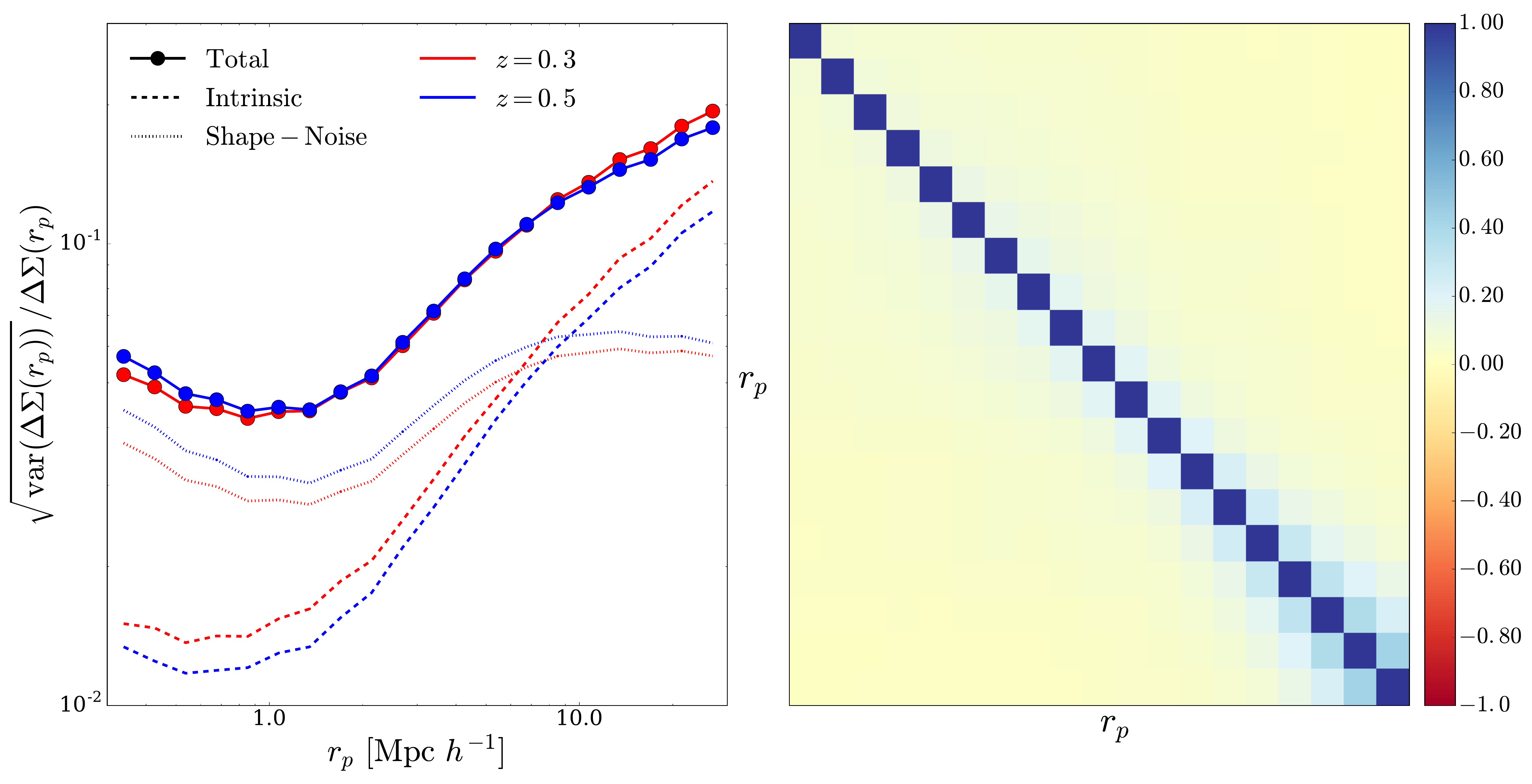}
\caption{Diagonal elements (left) of our adopted weak lensing covariance matrix for the $z = 0.3$ and $0.5$ cluster bins, converted to fractional errors, separately showing the intrinsic and shape noise contributions. Correlation matrix (right) of the $\Delta \Sigma$ observable at $z = 0.5$.}
\label{fig:delsig_cov}
\end{figure*}

\subsection{Numerical}
\label{sec:numerical}

To numerically estimate a measurement covariance matrix we use subvolumes of our 20 realizations of the fiducial cosmology. Each realization is subdivided into 25 equal volume regions by tiling a face of the box. The corresponding subvolumes are rectangular prisms, where the major axis is taken to be the line of sight.\footnote{This way we can satisfy the need to have the transverse size of the volume be significantly larger than $r_{p,\mathrm{max}}$, and likewise have the depth of the volume be significantly larger than $\pimax$.} In each subvolume we compute the observables and include pairs that cross the subvolume boundaries weighted by $0.5$. \citet{Friedrich_et_al_2016} have shown that discounting boundary pairs will artificially increase the variance due to removing the information these pairs provide. Conversely, including the cross-boundary pairs without weighting will artifically reduce the variance by duplicating pairs in adjacent subvolumes.

To compute the covariance we use a bootstrap method \citep[e.g.][]{Norberg_et_al_2009}. We sample 500 times with replacement from our $N_{\mathrm{sub}} = 500$ subvolumes and average the result to define a bootstrap resample:

\beq
\hat{\mathcal{O}_i} = \frac{1}{N_{\mathrm{sub}}} \sum_{j = 0}^{N_{\mathrm{sub}}} \mathcal{O}_{R_j^i},
\eeq
where $R_j^i$ is the $j$-th element of the $i$-th random sampling of $\left[1, 2, \dots, N_{\mathrm{sub}}\right]$ with replacement. The observable covariance is then calculated for a survey of volume $V_s$ by

\beq
\mathrm{cov}_{\mathcal{O}} \left(r_i, r_j \right) = \frac{V_{\mathrm{sub}}}{V_s} \sum_{i,j = 0}^{N_{\mathrm{sub}}} \left( \hat{\mathcal{O}}_i - \left< \hat{\mathcal{O}} \right> \right) \left( \hat{\mathcal{O}}_j - \left< \hat{\mathcal{O}} \right> \right) \label{eq:bootstrap}, 
\eeq
where $V_{\mathrm{sub}}$ is the volume of the individual subvolumes used to measure the covariance \footnote{If we had used a number of bootstrap samples $N_{\mathrm{samp.}} \neq N_{\mathrm{sub}}$, then the r.h.s of equation \ref{eq:bootstrap} would include an additional factor of $N_{\mathrm{sub}} / N_{\mathrm{samp.}}$.}. Figure \ref{fig:diag_comp} compares the diagonal elements of the covariance matrix - more precisely the fractional error $\sqrt{\mathrm{cov}_{\mathcal{O}} (r_i, r_i) } / \mathcal{O}_i$ - from our numerical and analytical estimates for the $z = 0.5$ cluster bin. At $r_p > 10.0 \; \Mpch$ there is good agreement of the two estimates, which is reassuring evidence that we have implemented both methods correctly. At $r_p < 10.0 \; \Mpch$ the numerical covariances are larger, as expected from the non-Gaussianity of clustering in the non-linear regime. This non-Gaussian contribution is larger for $w_{p,cg}$ than for $w_{p,gg}$

Figure \ref{fig:corr_matrix} compares our numerical and analytic estimates of the correlation matrix:

\beq
\mathrm{corr}_{\mathcal{O}}(r_{i},r_{j}) = \frac{\mathrm{cov}_{\mathcal{O}}(r_{i}, r_{j})}{\sqrt{\mathrm{cov}_{\mathcal{O}}(r_{i}, r_{i}) \mathrm{cov}_{\mathcal{O}}(r_{j}, r_{j})}}.
\eeq
In both cases the strongest off-diagonal elements are at large scales in $w_{p,gg}$, and cross-observable covariance is much smaller than the covariance within each observable. The numerical estimates shows correlations further from the diagonal than the analytic estimate, which is a plausible consequence of non-Gaussianity. However, the numerical correlation matrix is inherently noisy, and a noisy covariance matrix can artificially bias forecasts of parameter constraints to be too optimistic. We have therefore elected to use our numerical estimates of the diagonal errors but compute off-diagonal covariance by multiplying the analytic correlation matrix by these numerical diagonal elements. We show in section \cref{sec:forecasts} that our results would not change substantially if we were to use the numerically estimated covariance matrix or to ignore off-diagonal covariances entirely.

Figure \ref{fig:delsig_cov} shows the fractional errors from the diagonal elements of the $\Delta \Sigma$ covariance matrix at $z = 0.3$ and $z = 0.5$ and the correlation matrix of the $\Delta \Sigma$ errors at $z = 0.5$. For details of this calculation we refer the reader to \citet{Wu_et_al_2019}. At scales $r_p < \; 5.0 \Mpch$, the covariance is dominated by shape noise. At large $r_p$ the dominant source of statistical error is cosmic shear from uncorrelated foreground and background structure.

\section{Cosmological Forecasts}
\label{sec:forecasts}

\subsection{Fisher Information and Forecasting}
\label{sec:fisher}

Following the standard approach to Fisher matrix forecasting \citep[e.g.][]{Tegmark_1997,DodelsonModernCosmology_2003,DarkEnergyTaskForce_Albrecht_2009}, we write the Fisher information matrix as:

\beq
F_{ij} = \sum_{m,n} \frac{\partial \mathcal{O}(r_n)}{\partial \theta_i} \mathrm{cov}^{-1}(r_m, r_n) \frac{\partial \mathcal{O}(r_m)}{\partial \theta_j},
\eeq

where the derivatives of observables with respect to parameters are those derived in \cref{sec:derivs} and the covariance matrix is that derived in \cref{sec:cov}. Our forecast of the statistical error on a model parameter $\theta_i$ is $\left[ (F^{-1})_{ii} \right]^{1/2}$, and our estimate of the covariance for two parameters $\theta_i$, $\theta_j$ is $(F^{-1})_{ij}$.

\subsection{Fiducial scenario}
\label{sec:constraints}

\begin{figure*}
\centering
\includegraphics[width=1.0\textwidth]{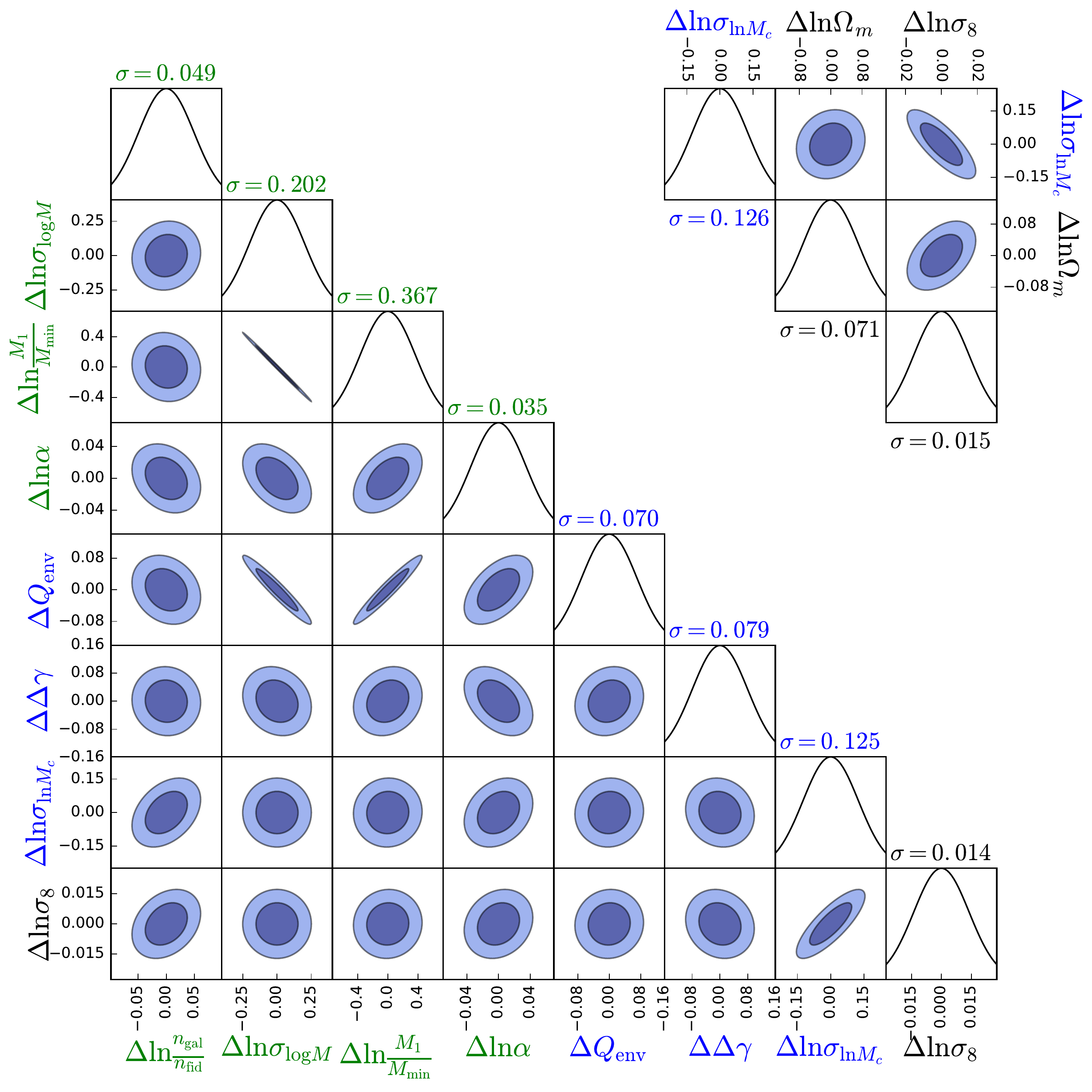}
\caption{Forecast parameter constraints ($68\%$ and $95\%$ contours) for our fiducial scenario, assuming a DES-like cluster and weak lensing sample, a cluster redshift bin $z = 0.35 - 0.55$, and using all scales $0.3 < r_p < 30.0 \; \Mpch$ in the $\Delta \Sigma$, $w_{p,cg}$, and $w_{p,gg}$ data vector. For the main block we hold $\Omega_m$ fixed at its fiducial value. The upper right block shows constraints on cosmological parameters and $\siglnMc$ when $\Omega_m$ is allowed to vary at fixed $\Omega_m h^2$. Fully marginalized errors on each parameter are listed above each PDF panel.}
\label{fig:fid_forecast}
\end{figure*}

\begin{table*}
   \centering
   \resizebox{\textwidth}{!}{\begin{tabular}{lllcccccccccc}
      \hline
        $\Delta \Sigma $ & $w_{p,cg}$ & $w_{p,gg}$ & $\Delta \ln \frac{\ngal}{\nfid}$ & $\Delta \ln \sigma_{\log M}$ & $\Delta \ln \frac{M_1}{M_{\mathrm{min}}}$ & $\Delta \ln \alpha$ & $\Delta Q_{env}$ & $\Delta \Delta \gamma$ & $\Delta \ln \sigma_{\ln M_c}$ & $\Delta \ln \sigma_8$\\
 	\hline
        all & all & all & 0.049 & 0.202 & 0.367 & 0.035 & 0.070 & 0.079 & 0.125 & 0.014 \\
        \hline
        all & - & - & . & . & . & . & . & . & 0.926 & 0.083 \\
        - & all & - & 0.050 & 3.257 & 4.531 & 0.734 & 0.293 & 0.162 & 5.818 & 0.152 \\
        - & - & all & 0.050 & 0.387 & 0.694 & 0.087 & 0.124 & 0.366 & . & 0.116\\
        \hline        
        - & all & all & 0.049 & 0.202 & 0.373 & 0.038 & 0.071 & 0.097 & 0.126 & 0.063 \\
        all & - & all & 0.050 & 0.382 & 0.694 & 0.078 & 0.124 & 0.366 & 0.755 & 0.068 \\
        all & all & - & 0.050 & 0.800 & 1.616 & 0.458 & 0.189 & 0.150 & 0.813 & 0.073 \\
        \hline
        large & large & large & 0.050 & 1.504 & 7.422 & 3.024 & 0.139 & 9.512 & 0.422 & 0.037\\
        large & large & all & 0.050 & 0.361 & 0.636 & 0.073 & 0.115 & 0.360 & 0.290 & 0.028\\
        small & small & small & 0.050 & 0.328 & 0.601 & 0.050 & 0.119 & 0.081 & 0.169 & 0.018\\
        small & small & all & 0.050 & 0.249 & 0.455 & 0.039 & 0.085 & 0.080 & 0.145 & 0.016\\
        \hline
         small & all & all & 0.049 & 0.202 & 0.367 & 0.035 & 0.070 & 0.079 & 0.125 & 0.014\\
        all & small & all & 0.050 & 0.249 & 0.455 & 0.039 & 0.085 & 0.080 & 0.143 & 0.015\\
        all & all & small & 0.050 & 0.242 & 0.441 & 0.038 & 0.083 & 0.080 & 0.130 & 0.014\\
        large & all & all & 0.049 & 0.202 & 0.367 & 0.035 & 0.070 & 0.080 & 0.125 & 0.018\\
        all & large & all & 0.050 & 0.356 & 0.629 & 0.073 & 0.113 & 0.359 & 0.283 & 0.026\\
        all & all & large & 0.050 & 0.427 & 0.978 & 0.306 & 0.118 & 0.134 & 0.312 & 0.029\\
        \hline
   \end{tabular}}
   \caption{Parameter forecast uncertainties with $\Omega_m$ fixed. Note that $\Delta \Sigma$ provides no information on HOD parameters and $w_{p,gg}$ provides no information on $\siglnMc$. 
   }
   \label{table:uncertainties}
\end{table*}

We forecast parameter constraints for our fiducial scenario, a DES-like survey, with the mixed numerical/analytic covariance matrix described in \cref{sec:cov} for $\Delta \Sigma$, $w_{p,cg}$, and $w_{p,gg}$. Derivatives are calculated directly and smoothed as described in \cref{sec:derivs}, and we additionally impose a $5\%$ Gaussian prior on the galaxy number density. Note that forecast parameters are in terms of the natural logarithm of the parameter of interest, except for parameters than can plausibly achieve zero or negative values such as $\Qenv$ and $\delgam$. With information from $0.3 \; \Mpch < r_p < 30.0 \; \Mpch$, a combination of $w_{p,cg}$, $\Delta \Sigma$ and $w_{p,gg}$ yields constraints on cosmology that are competitive with those from cosmic shear using the same weak lensing data set.

Figure \ref{fig:fid_forecast} and the top line of table \ref{table:uncertainties} present results for our ``fiducial'' case, using the full range of $r_p = 0.3 - 30.0 \; \Mpch$ for all three observables. To simplify interpretation, we consider \emph{only} the $z = 0.35 - 0.55$ redshift bin so that there is a single set of HOD parameters and a single value of $\siglnMc$ to constrain along with the cosmological parameters. We examine the gains from a second redshift bin in \cref{subsec:Cdef_Bin} below.

If we leave both $\sigma_8$ and $\Omega_m$ as free cosmological parameters, then the best constrained combination in our fiducial forecast is $\sigma_8 \Omega_m^{0.096}$, with a $1 \sigma$ uncertainty of $1.39\%$ after marginalizing over $\siglnMc$ and HOD parameters (top right portion of \cref{fig:fid_forecast}). The shallow slope of the degeneracy is a direct consequence of the relative insensitivity of our observables to $\Omega_m$ at fixed $\Omega_m h^2$, as seen in figures \ref{fig:derivatives}-\ref{fig:gg_frac_diffs}. As expected from this shallow slope, the marginalized uncertainty on $\Omega_m$ is much larger than that on $\sigma_8$, $7.07\%$ vs $1.54\%$. To further simplify our discussion we hereafter hold $\Omega_m$ fixed at its fiducial value and consider $\sigma_8$ as the sole cosmological parameter to be constrained (main body of figure \ref{fig:fid_forecast} and all rows of table \ref{table:uncertainties}). The fiducial forecast constraint on $\sigma_8$ is then $1.39\%$, a fractional error similar to that on $\sigma_8 \Omega_m^{0.096}$ when $\Omega_m$ is left free. 

This result is fairly robust with respect to our choice of covariance matrix. If we forecast with the numerical clustering covariance matrix, then the $\sigma_8$ constraint widens slightly to $1.42\%$, while using the analytic clustering covariance matrix tightens the constraint to $1.22\%$. If we only use the diagonal errors of our mixed analytic/numerical covariance matrix, we forecast a $1.35\%$ constraint on $\sigma_8$.

The strongest effect on the uncertainly in $\sigma_8$ is degeneracy with $\siglnMc$ (figure \ref{fig:fid_forecast}, bottom right). This behavior is expected from figure \ref{fig:cg_frac_diffs}, as increasing $\sigma_8$ and $\siglnMc$ simultaneously has a nearly cancelling effect on $\Delta \Sigma (r_p)$ at all scales. Among HOD parameters, there is strong degeneracy between $\siglogM$ and $M_1/\Mmin$, which have qualitatively similar effects on $w_{p,cg}(r_p)$ and $w_{p,gg}(r_p)$ (see figure \ref{fig:cg_frac_diffs}). The individual constraints on these parameters are therefore weak. These parameters are also degenerate with $\Qenv$ because of its impact on the large scale galaxy bias, but $\Qenv$ itself is quite well determined, with an uncertainty of $0.070$. This result bodes well for future efforts to constrain galaxy assembly bias with DES data. The parameters $\alpha$ and $\Delta \gamma$ are also well constrained because they affect small scales much more strongly than large scales, and because changing $\alpha$ has opposite effects on $w_{p,cg}(r_p)$ and $w_{p,gg}(r_p)$. The forecast constraint on the galaxy number density $\ngal$ is dominated by our $5\%$ prior. Fortunately, $\ngal$ is directly observable and it is not strongly degenerate with $\sigma_8$. If we change the $\ngal$ prior from $5\%$ to $1\%$ or $10\%$, the the $\sigma_8$ uncertainty changes from $1.39\%$ to $1.33\%$ or $1.54\%$, respectively.

We forecast a constraint of $12.45\%$ on $\siglnMc$. This value is roughly consistent with recent attempts to constrain the cluster richness-mass relationship. \citet{Murata_et_al_2018} constrained the relation using cluster abundance and stacked weak lensing profiles in bins of richness from redMaPPer selected SDSS clusters from $0.10 < z < 0.33$. They considered a more complicated form of the cluster mass-observable by allowing the scatter to change with mass. They modeled the scatter as a linear function in mass and were able to obtain $\approx 10 \%$ level constraints on the offset in this linear relation. Since \citet{Murata_et_al_2018} were principally interested in constraining the mass-observable relation, they did not marginalize over cosmology and instead chose a fixed Planck-like cosmology for their study. If we similarly fix cosmology then we forecast a constraint of $5.0\%$ on $\siglnMc$ with the combination of $\Delta \Sigma$, $w_{p,cg}$, and $w_{p,gg}$, marginalized over HOD parameters. This result shows the ability of this data combination to tightly constrain mass-observable scatter, and thus test models of cluster physics, when the cosmology is assumed to be known independently.

\subsection{Relative contributions of observables and scales}
\label{subsec:obs_and_scales}

To better understand the origin of the fiducial constraints, we examine a variety of alternative scenarios in Table \ref{table:uncertainties} in which we omit one or two of the observables or restrict them to small ($r_p < 3.0 \; \Mpch$) or large ($r_p > 3.0 \; \Mpch$) scales. We break at $3.0 \; \Mpch$ as an approximate division between the virial regime and the quasi-linear regime, and because our data vectors have equal numbers of points above and below this scale. The precision is higher for small-$r_p$ data points, as shown in figures \ref{fig:diag_comp} and \ref{fig:delsig_cov}. In all of these scenarios we hold $\Omega_m$ fixed and treat $\sigma_8$ as the sole cosmological parameter.

The second line in table \ref{table:uncertainties} shows our forecast for $\Delta \Sigma (r_p)$ as the only observable. The precision on $\sigma_8$ is drastically worse than the fiducial case, $\Delta \ln \sigma_8 = 0.083$ vs. $0.014$, because of the strong degeneracy between $\sigma_8$ and $\siglnMc$. These parameters do not have identical effects on $\Delta \Sigma (r_p)$ as a function of scale (see figure \ref{fig:cg_frac_diffs}, top row), so this degeneracy is weakly broken, but cluster weak lensing does not give competitive $\sigma_8$ constraints on its own unless there is an external prior on $\siglnMc$. We see from the next two lines that neither $w_{p,cg}$ nor $w_{p,gg}$ gives interesting $\sigma_8$ constraints on its own, with $\Delta \ln \sigma_8 > 0.1$ in each case. The $w_{p,cg}$ observable does not provide good HOD constraints, and inspection of figure \ref{fig:cg_frac_diffs} and table \ref{table:uncertainties} suggests this is a consequence of degeneracy between $\siglogM$ and $\siglnMc$, and between combinations of these parameters and $M_1 / \Mmin$. The $w_{p,gg}$ observable is unaffected by $\siglnMc$, and it yields much tighter HOD constraints. The marginalized errors on HOD parameters are still fairly large, however, perhaps because of 3-way degeneracy among $\siglogM$, $M_1 / \Mmin$, and $\Qenv$, as well as partial degeneracy with $\sigma_8$. \citet{Guo_2014} find much tighter constraints on the CMASS HOD parameters from BOSS galaxy clustering, but they assume a fixed cosmology and do not include an assembly bias parameter analogous to $\Qenv$.

The next three lines of table \ref{table:uncertainties} show forecasts for pairwise combinations of the three observables. The first key point is that none of these combinations yields a $\sigma_8$ constraint close to that of our own fiducial 3-observable combination; all have $\Delta \ln \sigma_8 > 0.06$ vs. $\Delta \ln \sigma_8 = 0.014$. Our tight fiducial constraint on $\sigma_8$ relies critically on the weak lensing observable, and supplementing $\Delta \Sigma$ with either $w_{p,cg}$ or $w_{p,gg}$ alone only slightly improves the $\sigma_8$ constraint relative to $\Delta \Sigma$ alone. However, the combination of $w_{p,cg}$ and $w_{p,gg}$ does yield a constrain on $\siglnMc$ that is nearly as good as that of the fiducial data combination, $12.6\%$ vs. $12.5\%$. This combination also yields much better HOD constraints then $w_{p,cg}$ or $w_{p,gg}$ alone, nearly as good as those from the full fiducial combination. These results support a fairly straightforward interpretation of the way the three observables interact to constrain $\sigma_8$. The two clustering observables jointly constrain HOD parameters and $\siglnMc$. The constraint on $\siglnMc$ in turn allows the weak lensing observable to cosntrain $\sigma_8$ instead of the degenerate combination of $\sigma_8$ and $\siglnMc$.

The remaining lines in table \ref{table:uncertainties} show the impact of restricting the data vector to small or large scales for one or more of the three observables. We first consider the case of using only the large scales in each observable. From Figures \ref{fig:cg_frac_diffs} and \ref{fig:gg_frac_diffs} one can see that for $r_p > 3.0 \; \Mpch$ all of the model parameters have a nearly scale-independent effect on the observables; to a good approximation they can be viewed as changing just the overall galaxy or cluster bias factor or (in the case of $\sigma_8$) the amplitude of $\xi_{mm}$. In the linear bias regime we expect

\begin{align}
\Delta \Sigma &\propto b_c \sigma_8^2, \label{eq:heuristic1}\\
w_{p,cg} &\propto b_c b_g \sigma_8^2, \label{eq:heuristic2}\\
w_{p,gg} &\propto b_g^2 \sigma_8^2, \label{eq:heuristic3}
\end{align}
so measurements of the three observables suffice to constrain the three unknowns $b_g$, $b_c$, and $\sigma_8$. With our adopted covariance matrices, the forecast error on $\sigma_8$ is $3.7\%$, about $2.5\times$ worse than the fiducial all-scales forecast, but substantially better than $\Delta \Sigma$ over all scales with no $\siglnMc$ prior. The errors on individual HOD parameters are very large because they are almost perfectly degenerate in this regime, but that degeneracy does not wreck the $\sigma_8$ constraint because the bias factor $b_g$ is constrained even if we do not know what HOD parameters lead to it. The constraint on the mass-observable scatter is $\Delta \ln \siglnMc = 0.422$, better than for $\Delta \Sigma$ alone, but in this case one should think of $\siglnMc$ as the ``trailing'' parameter: the observables directly constrain $b_c$ and $\sigma_8$, and $\siglnMc$ follows from these two parameters plus the cluster space density. Restoring small scales to the $w_{p,gg}$ data vector (the ``large large all'' line in table \ref{table:uncertainties}) produces much better constraints on the HOD parameters and significant improvement in the $\sigma_8$ constraint, from $3.7\%$ to $2.8\%$.

Using only the small scale data from the three observables yields $\Delta \ln \sigma_8 = 0.018$, a factor of two better than using only large scales and nearly as tight as the fiducial $\Delta \ln \sigma_8 = 0.014$. From Figures \ref{fig:cg_frac_diffs} and \ref{fig:gg_frac_diffs} we can see that small scales outperform large scales because the statistical errors per bin are smaller, the observables are more sensitive to the parameters, and scale-dependence can break parameter degeneracies. Restoring the large scales to $w_{p,gg}$ (the ``small small all'' line of table \ref{table:uncertainties}) produces marginal improvement in $\Delta \ln \sigma_8$, from $0.018$ to $0.016$. This case can be viewed as a generalization of the mass-to-number ratio method of \citet{Tinker_et_al_2012}. Instead of estimating mean cluster mass and galaxy counts, one takes $\Delta \Sigma (r_p)$ as a measure of cluster mass profiles out to virial scales, $w_{p,cg}(r_p)$ as a measure of number count profiles over the same range, and combines with galaxy HOD constraints from $w_{p,gg} (r_p)$ to infer $\sigma_8$.

The last six lines of table \ref{table:uncertainties} show forecasts that include all scales for two of the observables and small or large scales for the third. Using all scales for the clustering observables and only small scales for $\Delta \Sigma$ yields a result that is nearly the same as the all-scales fiducial forecast, with $\Delta \ln \sigma_8 = 0.014$. Trading small scale $\Delta \Sigma$ for large scale $\Delta \Sigma$ degrades the constraint moderately to $\Delta \ln \sigma_8 = 0.018$, because the statistical errors on $\Delta \Sigma$ are larger for the $r_p > 3.0 \; \Mpch$ data points than for the $r_p < 3.0 \; \Mpch$ data points. Since the small and large scales of $\Delta \Sigma$ independently yield good constraints on $\sigma_8$, it is initially surprising that using all scales in the fiducial forecast does not yield significant further improvement, i.e., $\Delta \ln \sigma_8 = (0.014^{-2} + 0.018^{-2})^{-1/2} = 0.011$ instead of $0.014$. However, for the ``all all all'' and ``small all all'' forecasts the precision on $\sigma_8$ is limited primarily by degeneracy with $\siglnMc$, and the constraint on $\siglnMc$ comes mainly for the clustering observables rather than weak lensing (see \cref{sec:conc} for further discussion). It is encouraging that, in combination with $w_{p,gg}$ and $w_{p,cg}$, the ``mass function regime'' and ``cluster-mass correlation regime'' of $\Delta \Sigma$ can separately yield good constraints on $\sigma_8$, allowing a cross-check of results at comparable precision. When systematic uncertainties such as cluster mis-centering or photo-z biases in cluster regions are added to the model via nuisance parameters, the combination of small and large scale $\Delta \Sigma$ measurements may help to mitigate their impact.

Turning to the remaining cases in table \ref{table:uncertainties}, we see that omitting large scale data for $w_{p,cg}$ or $w_{p,gg}$ alone produces negligible degradation for $\sigma_8$ and little degradation for HOD parameters. However, omitting the small scale data in either observable causes significant degradation, with $\Delta \ln \sigma_8 = 0.026-0.029$. This result demonstrates the importance of the HOD-based emulator approach developed here, which enables use of galaxy clustering observables into the fully non-linear regime.

\begin{figure}
\centering
\includegraphics[width=0.45\textwidth]{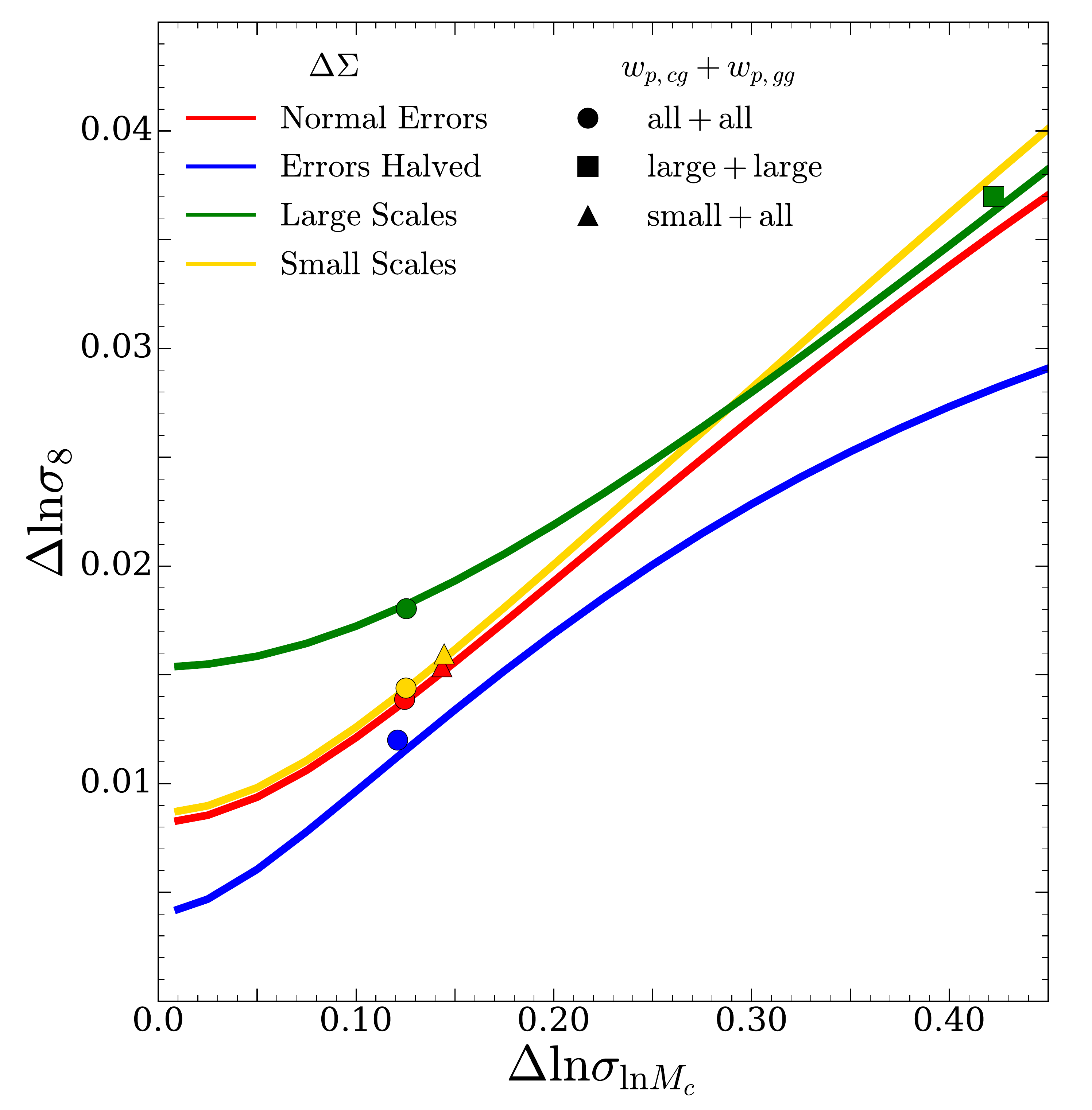}
\caption{Constraints from lensing only and a prior on $\siglnMc$ as a function of that external prior on $\siglnMc$. The red line shows the constraints using our fiducial covariance matrix, while the blue shows the same with the errors halved. Points are plotted in the $\Delta \siglnMc$-$\Delta \sigma_8$ locations that their respective scenarios (as listed in table \ref{table:uncertainties}) forecast with labels referring to the scales used of $w_{p,cg}$ and $w_{p,gg}$. For instance the red circle is plotted at the locations of the $\sigma_8$ and $\siglnMc$ constraints for our fiducial scenario, and the filled blue circle shows the corresponding constraints if the $\Delta \Sigma$ errors are reduced by a factor of two. The green square represents the linear theory limit of equations \ref{eq:heuristic1} - \ref{eq:heuristic3}, and the yellow triangle is analagous to the mass-to-number method method of \citet{Tinker_et_al_2012}. The open blue circle shows the constraint after errors on all three observables are reduced by a factor of two.}
\label{fig:siglnMc_priors}
\end{figure}

Figure \ref{fig:siglnMc_priors} summarizes some of the key results from Table \ref{table:uncertainties} in graphical form. The red curve shows the constraint on $\sigma_8$ that could be obtained from $\Delta \Sigma (r_p)$ and an external prior on $\siglnMc$. For a perfect prior ($ \Delta \ln \siglnMc = 0$) this constraint is $\Delta \ln \sigma_8 = 0.0081$, limited purely by the statistical uncertainties in the $\Delta \Sigma$ measurements. Green and yellow curves show the corresponding constraints using only the large or small scales of $\Delta \Sigma$, respectively. Constraints from small scales are nearly as good as those from all scales, while constraints from large scales are about a factor of two weaker if $\siglnMc$ is known. The blue curve shows the constraint on $\sigma_8$ if the $\Delta \Sigma$ errors are reduced by a factor of two. 

Points show our forecast constraints on $\sigma_8$ and $\siglnMc$ for the three-observable combination, using all scales of $w_{p,cg}$ and $w_{p,gg}$ (circles), using only large scales of these observables (squares), or using all scales of $w_{p,gg}$ and small scales of $w_{p,cg}$ (triangles). The point color indicates which scales of $\Delta \Sigma$ are used. The red circle represents our fiducial case, with $\Delta \ln \sigma_8 = 0.014$ and $\Delta \ln \siglnMc = 0.125$. The green square represents the linear regime limit (``large large large''), which gives substantially weaker constraints. The yellow triangle represents the ``mass-to-number ratio'' analogue (``small small all''), which is nearly as constraining as the use of all scales of all observables. Using all scales with a factor-of-two improvement in $\Delta \Sigma$ errors (blue circle) sharpens the $\sigma_8$ constraint from $1.39\%$ to $1.16\%$, much less than a factor of two because the clustering constraint on $\siglnMc$ has not improved. Exploiting the $\Delta \Sigma$ improvement requires smaller errors  in the clustering observables, as illustrated by the open blue circle, which has halved errors for all three observables and $\Delta \ln \sigma_8 = 0.007$. In all cases points lie close to the corresponding colored curve, and the uncertainty in $\sigma_8$ is significantly larger than it would be if $\siglnMc$ were perfectly known.

\subsection{Multiple mass and redshift bins}
\label{subsec:Cdef_Bin}

Our forecasts above are for a cluster sample corresponding to a mass threshold $M_c \approx 2 \times 10^{14} \Msun$. More precisely, we apply a threshold in $M_{\mathrm{obs}}$ to select a sample with space density $n_c = 3.254 \times 10^{-6} \invMpchCubed$ that equals the mean space density of halos with mass $M \geq 2 \times 10^{14} M_{\odot}$ in the fiducial cosmology at $z = 0.5$. In DES, clusters of this mass have a redMaPPeR richness $\lambda \geq 30$ \citep{McClintock_DES_2019}, high enough for robust selection. It may be feasible to lower the effective mass threshold to $1 \times 10^{14} \; \Msun$ and still select clusters and measure their richness with adequate signal-to-noise ratio. This boosts the cluster space density by a factor of $\sim 4.3$, enabling higher precision measurements of $w_{p,cg}$ and, more importantly, of $\Delta \Sigma$. If we repeat our fiducial forecast with this lower $M_{\mathrm{obs}}$ threshold we obtain a marginalized constraint on $\sigma_8$ of $0.9\%$ instead of $1.4\%$. Conversely, if we adopt an $M_{\mathrm{obs}}$ threshold corresponding to $M \geq 4 \times 10^{14} M_{\odot}$ in our fiducial cosmology, then the cluster space density is lower by a factor of $\sim 6.4$, and our forecast constraint on $\sigma_8$ from combining $\Delta \Sigma$, $w_{p,cg}$, and $w_{p,gg}$ loosens to $3.3\%$. 

Table \ref{table:redshifts&clusterdefs} lists the forecast constraints on $\sigma_8$ and $\siglnMc$ for these three mass thresholds, for the $z = 0.35 - 0.55$ redshift bin and for the $z = 0.15 - 0.35$ redshift bin. Because the comoving volume of the low-$z$ bin is a factor 1.8 smaller, the number of clusters is smaller, and the errors on $w_{p,cg}$ and $\Delta \Sigma$ are  therefore larger. The resulting errors on $\sigma_8$ are a factor $\sim 1.4-1.8$ larger than those for the $z = 0.35 - 0.55$ bin. Measurements for the two redshift bins should be essentially independent, so the errors derived from combining the two redshift bins are smaller. Note that we allow independent HOD parameters {\emph{and}} $\siglnMc$ values for the two redshift bins. Although the results for the ``both'' rows in Table \ref{table:redshifts&clusterdefs} are derived from a full Fisher forecast (with no measurement covariance between the redshift bins), they follow a simple quadrature combination of errors from the two bins, i.e., $\left( \Delta \ln \sigma_8 \right)_{\mathrm{both}} = \left[ (\Delta \ln \sigma_8)^{-2}_{\mathrm{high-z}} + (\Delta \ln \sigma_8)^{-2}_{\mathrm{low-z}}\right]^{-1/2}$.

One might imagine combining multiple mass thresholds at a given redshift could yield even tighter constraints. An advantage relative to combining the same mass threshold in different redshift bins is that one need only model one set of galaxy HOD parameters (but still allowing separate $\siglnMc$ values at each threshold), since the same galaxy population is used for all three $w_{p,cg}$ measurements. The covariance of the measurement errors {\emph{between}} samples will not be zero since they share clusters and therefore shapes. Despite this, if the derivatives for different cluster samples are sufficiently different in a way that can break parameter degeneracies this {\emph{cross-observable}} covariance can be overcome to provide even tighter constraints than simply using the most abundant sample. If we treat the errors between the samples as independent we forecast $\sigma_8$ errors of $0.65\%$ at $z = 0.35-0.50$ and $0.94\%$ at $z = 0.15-0.35$. However when we include the cross-observable contributions to the covariance we find that combining mass thresholds yields constraints consistent with those that come from the most abundant sample alone (see Appendix \ref{appendix:cross_cov} for more details). We therefore conclude that there is no advantage to using multiple mass thresholds or bins and that one should simply choose the lowest threshold for which systematic uncertainties associated with cluster identification do not substantially degrade the statistical power.

\begin{figure}
\centering
\includegraphics[width=0.45\textwidth]{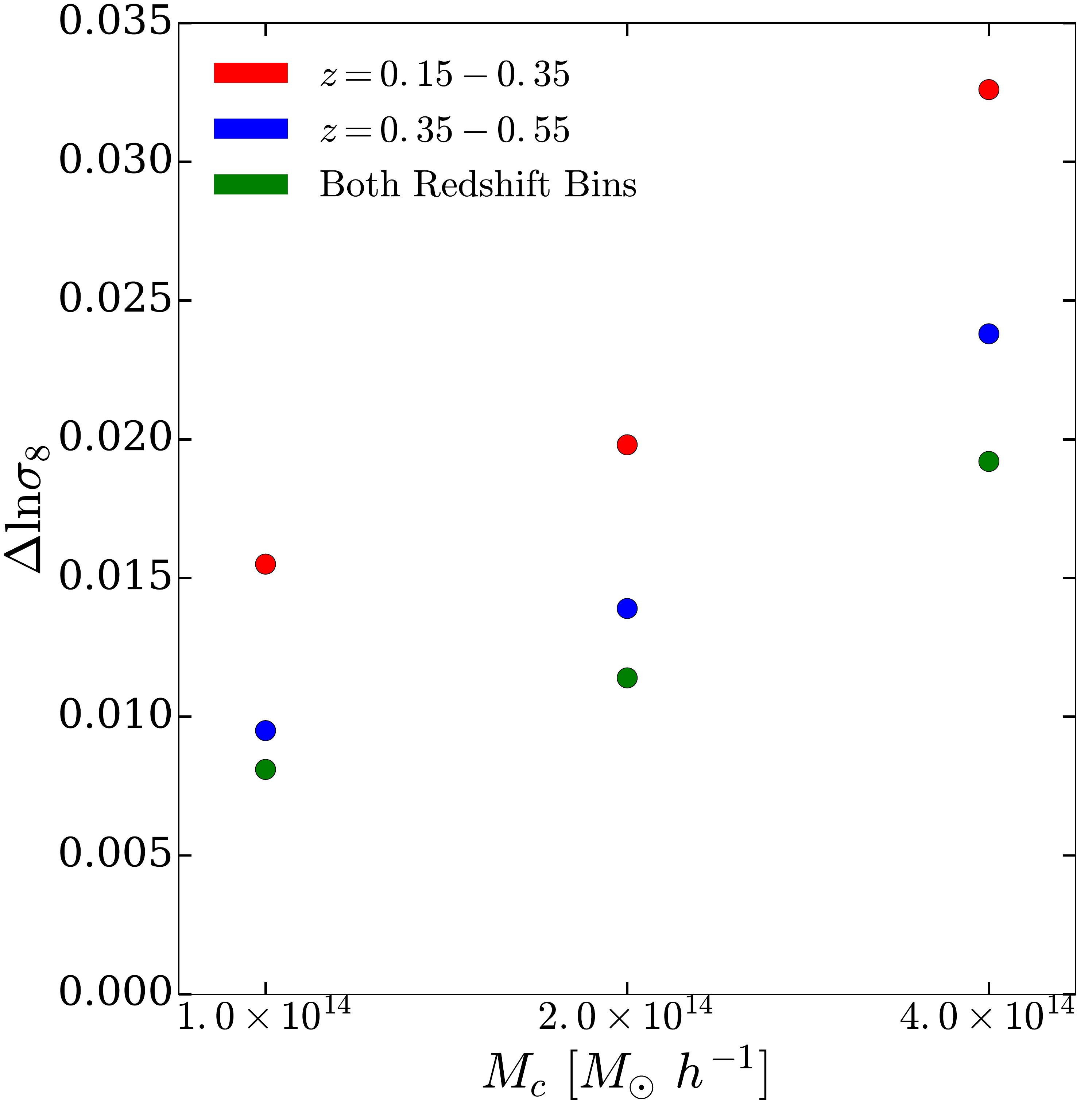}
\caption{Forecast constraints on $\ln \sigma_8$ from combinations of redshift bins and cluster samples. Red points show the constraints from our cluster samples at $z = 0.15 - 0.35$, and blue at $z = 0.35-0.55$, as a function of the $M_{\mathrm{obs}}$ threshold $M_c$. Green points show the combination of both redshift bins for each cluster definition.}
\label{fig:combined_fig}
\end{figure}

Figure \ref{fig:combined_fig} plots the $\sigma_8$ constraints from Table \ref{table:redshifts&clusterdefs} against the cluster mass threshold. If cluster identification remains reliable down to $10^{14} \; \Msun$ and systematics can be held sub-dominant to statistical errors, our forecasts imply that DES cluster measurements could obtain a sub-percent constraint on the amplitude of matter clustering at $z \approx 0.5$.

\subsection{Cluster number density uncertainties}
\label{subsec:nc_uncertainties}

We have evaluated the derivatives in \cref{sec:derivs} at fixed comoving cluster space density $n_c$, and our forecasts thus far implicitly assume that the space density of the cluster sample is known perfectly. In practice the value of $n_c$ has both statistical uncertainties from Poisson fluctuations and large scale structure and potential systematic uncertainties from incompleteness and contamination. For our $z = 0.35-0.55$ bin,  Figure 23c of WMEHRR implies statistical uncertainties of $1-2\%$ in $n_c$ for mass thresholds of $1-2 \times 10^{14} \; \Msun$, rising to $4\%$ for a mass threshold of $4 \times 10^{14} \; \Msun$. (Relative to the WMEHRR calculation, our fiducial scenario has half the survey area but double the $\Delta z$ bin width, approximately cancelling effects.)

\begin{figure}
\centering
\includegraphics[width=0.45\textwidth]{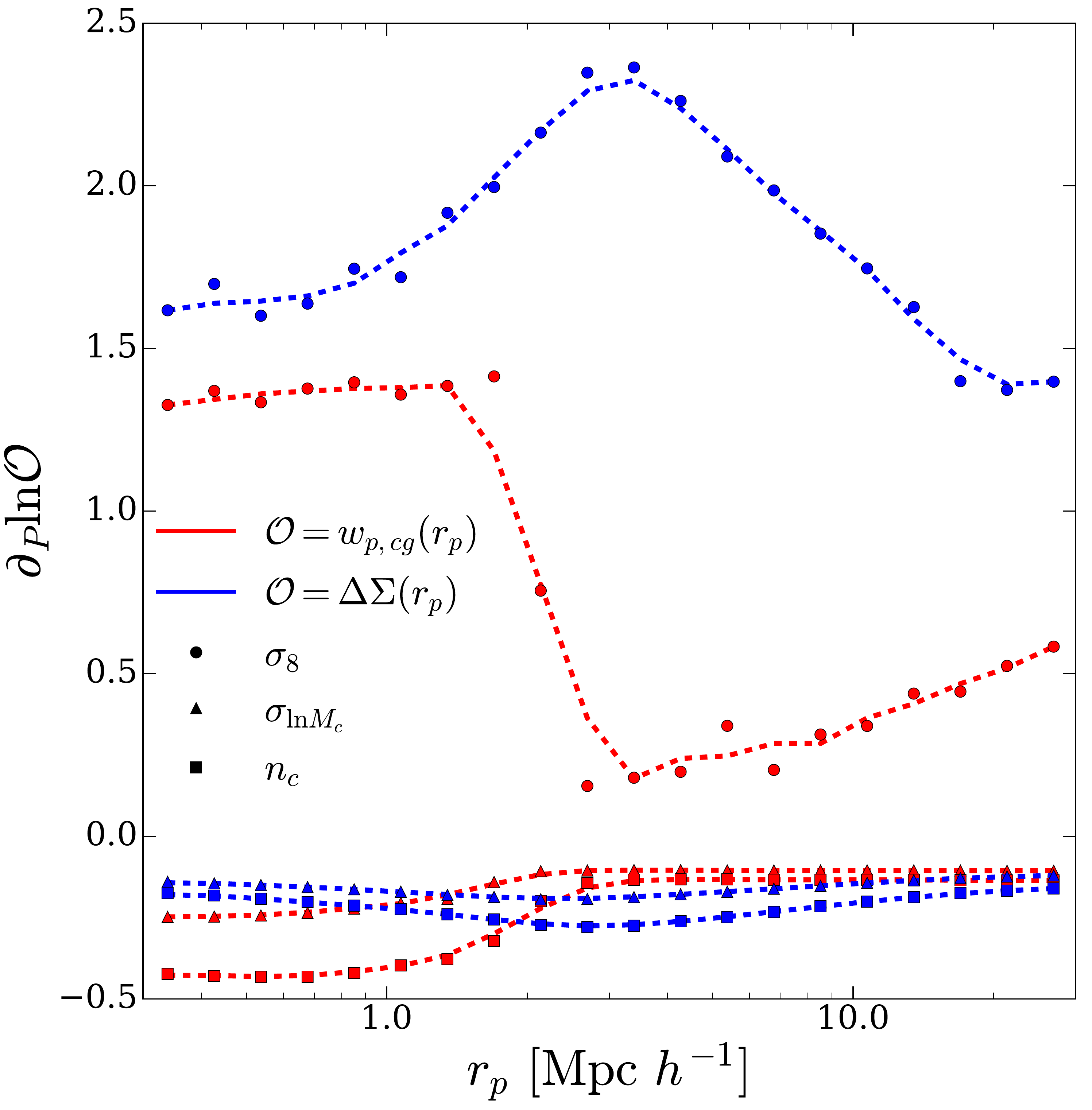}
\caption{Derivatives of $w_{p,cg}(r_p)$ (red curves and points) and $\Delta \Sigma (r_p)$ (blue curves and points) with respect to $\sigma_8$ (circles), $\siglnMc$ (triangles), and cluster mean space density $n_c$ (squares). Points show direct numerical estimates and curves show smoothed estimates.}
\label{fig:nc_deriv}
\end{figure}

To assess the impact of $n_c$ uncertainties, we have evaluated derivatives of $\Delta \Sigma$ and $w_{p,cg}$ with respect to $n_c$ by creating new cluster samples perturbed about our fiducial case. Figure \ref{fig:nc_deriv} shows that these derivatives are similar to the $\siglnMc$ derivatives, though with somewhat stronger sensitivity of $w_{p,cg}(r_p)$ at small scales $\left( r_p \leq 2 \; \Mpch \right)$. This similarity suggests that the cluster observables depend on a degenerate combination of $n_c$ and $\siglnMc$, with little sensitivity to the parameters individually.

First consider a case in which $\Delta \Sigma (r_p)$ and $n_c$ are the only observables and $\siglnMc$ is known from an external prior. This is essentially a cumulative form of the traditional cluster mass function approach, with weak lensing mass calibration and a single mass threshold in place of multiple bins. If $n_c$ and $\siglnMc$ are perfectly known, then the $\Delta \Sigma (r_p)$ covariance for our $z = 0.35-0.55$ cluster bin leads to an error on $\sigma_8$ of $0.8\%$ (the $\Delta \siglnMc = 0$ limit of the red curve in Figure \ref{fig:siglnMc_priors}). Adding a $2\%$, $5\%$, or $10\%$ error on $n_c$, while retaining perfect knowledge of $\siglnMc$, increases the $\sigma_8$ error to $0.86\%$, $1.03\%$, $1.48\%$, repectively. If we instead impose a $\siglnMc$ prior of $12.5\%$, equal to the value of $\Delta \siglnMc$ obtained from our fiducial combination of $w_{p,cg}$ and $w_{p,gg}$, then the $\sigma_8$ errors are $1.38\%$, $1.40\%$, $1.51\%$, and $1.84\%$ for $n_c$ uncertainties of zero, $2\%$, $5\%$, $10\%$, respectively. Errors on $n_c$ equal to the expected statistical error in a DES-like survey therefore add negligibly to the forecast error on $\sigma_8$, but $10\%$ errors (e.g., from completeness uncertainty) would noticeably degrade the $\sigma_8$ constraint.
 
Now consider the fiducial three-observable combination, with a wide prior on $\siglnMc$. Errors on $n_c$ of zero, $2\%$, $5\%$, and $10\%$ yield $\sigma_8$ uncertainties of $1.39\%$, $1.39\%$, $1.39\%$, and $1.41\%$. The $\sigma_8$ error for the three-observable combination thus degrades much more slowly with $n_c$ uncertainty than it does for the $\Delta \Sigma (r_p)$ case with a $\siglnMc$ prior. The error on $\siglnMc$ {\emph{does}} degrade, from $12.5\%$ to $12.9\%$, $15.1\%$, and $20.7\%$, but that is because $n_c$ and $\siglnMc$ have approximately degenerate effects on observables and only trade off against $\sigma_8$ in a combination that remains well determined. Even a $20\%$ uncertainty in $n_c$ only degrades the $\sigma_8$ error to $1.47\%$. 

The insensitivity to $n_c$ uncertainty highlights the fact that combining the cluster weak lensing with the galaxy clustering observables is {\emph{not}} closely analagous to measuring the cluster mass function, in which case $n_c$ uncertainties at the $10-20\%$ level would matter relative to DES-like errors in the weak lensing mass scale. Instead, as discussed in \cref{subsec:obs_and_scales}, the combination on large scales is roughly equivalent to using three observables to measure $\sigma_8$, $b_c$, and $b_g$, while on small scales it is roughly equivalent to the mass-to-number ratio method of \citet{Tinker_et_al_2012}. In both of these approaches knowledge of the cluster space density is not necessary for deriving cosmological constraints.

\begin{table*}
   \centering
   \begin{tabular}{ccccc}
      \hline
        $M_c \; \left[ h^{-1} \; \Msun \right]$ & $n_c \; \left[ \invMpchCubed \right]$ & $z$ & $\Delta \ln \siglnMc$ & $\Delta \ln \sigma_8$ \\
 	    \hline
 	    $1.0 \times 10^{14}$ & $2.13 \times 10^{-5}$ & $[0.15, 0.30]$ & 0.1906 & 0.0155 \\
 	    $2.0 \times 10^{14}$ & $5.85 \times 10^{-6}$ & $[0.15, 0.30]$ & 0.2060 & 0.0198 \\
 	    $4.0 \times 10^{14}$ & $1.15 \times 10^{-6}$& $[0.15, 0.30]$ & 0.2749 & 0.0326 \\
 	    $1.0 \times 10^{14}$ & $1.40 \times 10^{-5}$& $[0.35, 0.55]$ & 0.0996 & 0.0095 \\
 	    $2.0 \times 10^{14}$ & $3.25 \times 10^{-6}$& $[0.35, 0.55]$ & 0.1245 & 0.0139 \\
 	    $4.0 \times 10^{14}$ & $5.05 \times 10^{-7}$& $[0.35, 0.55]$ & 0.1690 & 0.0238 \\
 	    $1.0 \times 10^{14}$ & . & both & . & 0.0081 \\
 	    $2.0 \times 10^{14}$ & . & both & . & 0.0114 \\
 	    $4.0 \times 10^{14}$ & . & both & . & 0.0192 \\
    \hline
   \end{tabular}
   \caption{Parameter forecast uncertainties from combining different cluster samples and redshift bins with $\Omega_m$ fixed.}
   \label{table:redshifts&clusterdefs}
\end{table*}

\section{Conclusions}
\label{sec:conc}

We have investigated the cosmological constraints that can be obtained by combining mean cluster weak lensing profiles $\Delta \Sigma (r_p)$ with projected cluster-galaxy cross correlations $w_{p,cg}(r_p)$ and galaxy auto-correlations $w_{p,gg}(r_p)$. We compute observables as a function of model parameters using N-body simulations from the {\sc{abacus}} cosmology suite \citep{Garrison_et_al_2018}, populating dark matter haloes with galaxies using an HOD parameterization that includes an environmental dependence $\Qenv$ to represent possible effects of galaxy assembly bias. For our fiducial Fisher matrix forecasts we assume a DES-like survey of clusters and galaxies, focusing on the redshift range $z = 0.35 - 0.55$. We assume that DES can identify clusters above a halo mass threshold $M_c \approx 2 \times 10^{14} \; \Msun$ and that the relation between true halo mass and observable richness $\lambda$ (or other observable mass indicator) is log-normal with scatter $\siglnMc$. We choose $\siglnMc = 0.4$ for our fiducial value, which is somewhat pessimistic relative to empirical estimates (e.g., $\siglnMc \sim 0.25$ from \citet{Rozo_et_al_2014}), and we adopt a wide prior on $\siglnMc$ when computing our 3-observable forecasts. We assume fiducial galaxy HOD parameters similar to those inferred observationally for BOSS CMASS galaxies \citep{Guo_2014}. Our fiducial parameter choices and survey assumptions are detailed in Tables \ref{table:params} and \ref{table:covariance}. We predict covariance matrices of our observables using a mixture of numerical and analytic methods as described in \cref{sec:cov}. 

Our fiducial forecast, using all three observables over the range $r_p = 0.3 - 30.0 \; \Mpch$, yields a $1.4\%$ constraint on $\sigma_8$ for fixed $\Omega_m$, and a $12.5\%$ constraint on $\siglnMc$. If we leave $\Omega_m$ free (but $\Omega_m h^2$ fixed), then the best constrained parameter combination is approximately $\sigma_8 \Omega^{0.1}$, but for clarity of interpretation we adopt fixed $\Omega_m$ for most of our forecasts. If we omit the $\Delta \Sigma (r_p)$ observable then the $\siglnMc$ constraint is almost unchanged, but the $\sigma_8$ constraint degrades drastically to $6.8\%$; not surprisingly, the weak lensing data are crucial to constraining the dark matter clustering. Conversely, if we omit the $w_{p,cg}$ and $w_{p,gg}$ observables then the $\sigma_8$ constraint degrades drastically to $8.3\%$ because of the strong degeneracy between $\sigma_8$ and $\siglnMc$. We can thus interpret the fiducial forecast as follows: the two galaxy clustering observables together constrain $\siglnMc$, and this constraint allows $\Delta \Sigma$ to constrain $\sigma_8$. The alternative combinations investigated in Table \ref{table:uncertainties} show that both clustering observables are needed for this combination to work. On its own, $w_{p,gg}$ contains no information about $\siglnMc$, and $w_{p,cg}$ alone yields poor constraints on $\siglnMc$ because of degeneracy with HOD parameters.

Table \ref{table:uncertainties} also shows the effect of restricting one or more of the observables to small scales ($r_p = 0.3 - 3.0 \; \Mpch$) or large scales ($r_p = 3.0 - 30.0 \; \Mpch$). Using all scales of $w_{p,cg}$ and $w_{p,gg}$ but only the small scales of $\Delta \Sigma$ yields an equally strong $\sigma_8$ constraint of $1.4\%$. Using the large scales of $\Delta \Sigma$ instead of the small scales yields a constraint of $1.8\%$, somewhat weaker because of the larger observational errors on $\Delta \Sigma$ at large scales. It is encouraging that these two independent regimes of $\Delta \Sigma$ can both yield tight constraints on $\sigma_8$, allowing consistency checks and reducing sensitivity to any observational or theoretical systematics that would affect the small and large scales of $\Delta \Sigma$ differently. The cominbation of small scale $\Delta \Sigma$ and $w_{p,cg}$ with all scales of $w_{p,gg}$, which can be regarded as a correlation function form of the \citet{Tinker_et_al_2012} mass-to-number ratio method, yields a $\sigma_8$ constraint of $1.5\%$. Restricting all three observables to large scales, where perturbation theory with bias factors may provide an adequate description, produces a substantially weaker $\sigma_8$ constraint of $3.7\%$.

The $z = 0.15 - 0.35$ cluster redshift bin has lower volume and therefore yields a weaker but still interesting $\sigma_8$ constraint of $2.0\%$. At either redshift, lowering or raising the cluster mass threshold (to $1.0 \times 10^{14} \; h^{-1} \; \Msun$  or $4 \times 10^{14} \; h^{-1} \Msun$) strengthens or weakens the $\sigma_8$ constraint, respectively. The minimum achievable mass threshold will be set in practice by the reliability of cluster identification and richness estimation. 

There are numerous potential systematics in an observational analysis that are not addressed in our idealized study. However there are reasons to think that the three-observable approach outlined here may be less sensitive to systematics than a traditional cluster mass function analysis. Our forecast constraints are insensitive to uncertainties in the cluster space density, which could arise from incompleteness or contamination (see \cref{subsec:nc_uncertainties}). Baryonic physics including AGN feedback may alter cluster mass profiles on small scales, but our use of the full $\Delta \Sigma (r_p)$ profiles rather than mass within a specified radius should at least mitigate these effects, and those of cluster mis-centering. Systematic uncertainties in weak lensing shear calibration and photo-$z$ distributions are important for any cosmological constraints that use weak lensing, but the different sensitivities of cluster weak lensing and cosmic shear and of different scales of $\Delta \Sigma (r_p)$ may help to constrain nuisance parameters that describe these effects. Potentially the most difficult cluster-specific systematics arise from anisotropies in cluster identification and richness estimation, e.g., artificially boosting the richness of ellipsoidal clusters that are oriented along the line of sight or spuriously blending groups and clusters that are distinct in three dimensions but superposed in projection \citep{Constanzi_et_al_2019,Ramos-Ceja_et_al_2019}. The combination of $\Delta \Sigma (r_p)$ and $w_{p,cg} (r_p)$ may mitigate these systematics because the same effects that artificially boost $\Delta \Sigma$ will artificially boost $w_{p,cg}$, as suggested by \citet{Tinker_et_al_2012} in the context of the mass-to-number ratio method. For now this mitigation is simply a conjecture, which will need to be tested with simulations that mimic in detail the procedures for cluster identification, richness estimation, and measurement of cluster-galaxy cross-correlations. 

In DES the approach advocated here could be implemented using a galaxy sample with photometric selection that mimics that used for BOSS CMASS spectroscopic targets \citep[][]{Dawson_BOSS_et_al_2013, Lee_DMASS_2019}. Alternatively, it could be implemented with the redMaGiC galaxy sample \citep{Rozo_Rykoff_redmagic_et_al_2016}, which is designed to have high photo-z accuracy, though this will require a modified HOD formulation. This approach should also be applicable to the deeper imaging surveys expected from Subaru HSC, LSST, Euclid, and WFIRST, with either the optically identified clusters from these surveys or X-ray selected clusters from eROSITA. The forecasts of WMEHRR suggest that measurements of dark matter clustering from cluster weak lensing have comparable power to measurements from cosmic shear in the same weak lensing data set, adding significant leverage for distinguishing between dark energy and modified gravity explanations of cosmic acceleration (see, e.g., WMEHRR figures 46 and 47). Combining cluster weak lensing with cluster-galaxy cross-correlations and galaxy auto-correlations may prove the more robust route to realizing this promise.

\section*{Acknowledgements}

We thank Lehman Garrison, Chris Hirata, Ashley Ross, Ying Zu and the OSU-CCAPP cosmology group for valuable conversations about this work. ANS is supported by the Department of Energy Computational Science Graduate Fellowship Program of the Office of Science and National Nuclear Security Administration in the Department of Energy under contract DE-FG02-97ER25308. BDW is supported by the National Science Foundation Graduate Research Fellowship Program under Grant No. DGE-1343012. Any opinions, findings, and conclusions or recommendations expressed in this material are those of the author(s) and do not necessarily reflect the views of the National Science Foundation. This work was supported in part by NSF Grant AST-1516997 and NASA Grant 15-WFIRST15-0008. Simulations were analyzed in part on computational resources  of  the  Ohio  Supercomputer  Center \citep{OhioSupercomputerCenter1987},with resources supported in part by the Center for Cosmology and AstroParticle Physics at the Ohio State University. Some computations in this paper were performed on the El Gato supercomputer at the University of Arizona, supported by grant 1228509 from the National Science Foundation, and on the Odyssey cluster supported by the FAS Division of Science, Research Computing Group at Harvard University. We gratefully acknowledge the use of the {\sc{matplotlib}} software package \citep{Hunter_2007} and the GNU Scientific Library \citep{GSL_2009}. This research has made use of the SAO/NASA Astrophysics Data System.

\bibliographystyle{mnras}
\bibliography{masterbib}

\appendix

\section{Cross-Observable Covariance}
\label{appendix:cross_cov}
 
\begin{figure*}
\centering
\includegraphics[width=1.0\textwidth]{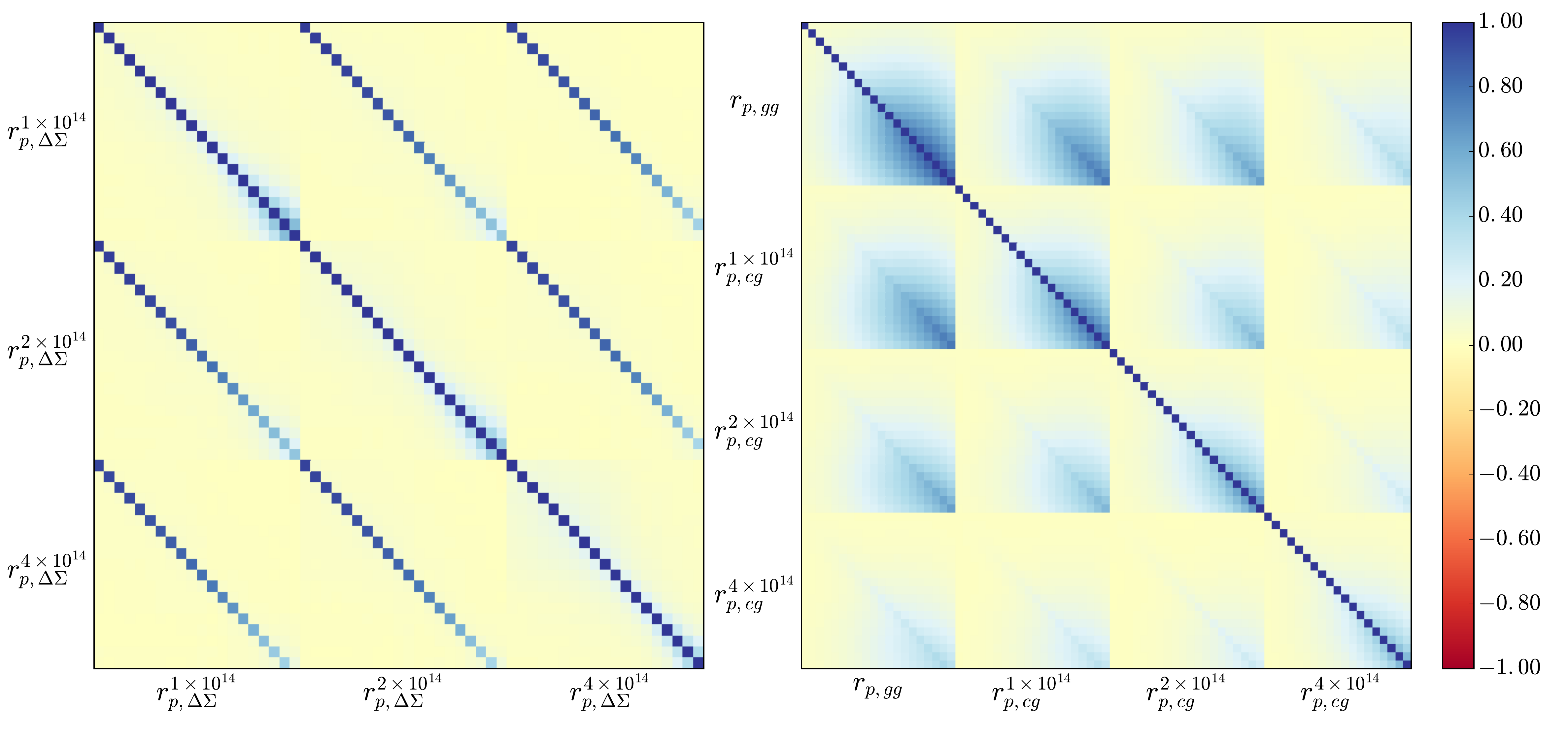}
\caption{Correlation matrix of $\Delta \Sigma$ (left) and $w_{p,gg}$ and $w_{p,cg}$ (right) for all cluster samples considered including off-diagonal cross-sample components.}
\label{fig:cross_corr}
\end{figure*}

 In Section \ref{subsec:Cdef_Bin} we discuss the possibility of combining observables from multiple cluster samples to obtain even tighter constraints on $\sigma_8$ than from any one sample alone. Since the most abundant cluster sample contains all the clusters in the other samples, it is unclear whether this combination will yield better constraints than those obtained from just using the most abundant sample. In principle this is possible, provided that the cosmological derivatives in different samples are sufficiently different to break degeneracies that exist in the derivatives of any given cluster sample. To do such a test properly we must include the covariance {\emph{between}} different cluster samples. Using equation \ref{eq:gen_cov} we compute the cross-observable covariance for $w_{p,cg}$ between two different cluster samples,
 
\begin{align}
\mathrm{cov} \left( w_{p,c_1g} \left( r_{p,i} \right), w_{p,c_2 g} \left(r_{p,j}\right) \right) = \frac{2\pimax}{V_s} \int_0^{\infty} \frac{k dk}{2\pi}  \hat{J}_0(k r_i)  \hat{J}_0(k r_j) \\
\times \left[ P_{c_1 c_2}(k)  \left( P_{gg}(k) + \frac{1}{n_g} \right) + P_{c_1 g}(k) P_{c_2 g}(k)\right]. \nonumber
\end{align}
The analogous expression for $\Delta \Sigma$ can be found in \citet{Wu_et_al_2019}. Since $\Delta \Sigma$ is shape-noise dominated in all of the scenarios we consider we ignore the covariance between $\Delta \Sigma$ and our clustering observables. 

Figure \ref{fig:cross_corr} shows the correlation matrices for $\Delta \Sigma$ (left) and $w_{p,cg}$ and $w_{p,gg}$ (right) including cross-observable contributions. In the case of lensing we see that the shape noise covariance between different samples is significant (because each cluster-source pair in the high mass threshold is also present in the lower mass threshold samples). For clustering we find a much weaker cross-observable component compared to the case of lensing.

To investigate the importance of the cross-observable component of the covariance, we forecast cases in which we combine $\Delta \Sigma$ and $w_{p,cg}$ from all three of our cluster samples with $w_{p,gg}$ in both of our redshift bins. We first forecast for this scenario with the cross-observable covariance set to zero. When we do so we forecast $\sigma_8$ errors of $0.65\%$ and $0.94\%$ in the $z \sim 0.5$ and $z \sim 0.3$ bins respectively. These are significantly better than a simple quadrature combination of results for the three mass bins because only one set of HOD parameters needs to be determined, not three. If we instead include the cross-observable covariance as shown in figure \ref{fig:cross_corr} we forecast $\sigma_8$ errors of $1.01\%$ and $1.60\%$ in the $z \sim 0.5$ and $z \sim 0.3$ bins respectively. These values are slightly large than the $\sigma_8$ error of $0.996\%$ from the most abundant ($M_c = 1 \times 10^{14} \; h^{-1} \; \Msun$) cluster sample on its own. This is because in the multiple cluster sample case $\sigma_8$ is marginalized over a $\siglnMc$ parameter for each cluster sample. The fact that the $\sigma_8$ error actually gets worse with additional information could be a consequence of marginalizing over more nuisance parameters (two additional values of $\siglnMc$), though the small change could also be affected by slight inaccuracies in our cross-observable covariance. However, it seems safe to conclude that using multiple mass thresholds does not improve $\sigma_8$ constraints.

\label{lastpage}

\end{document}